\documentclass[%
 reprint, pre,
superscriptaddress,
 amsmath,amssymb,
 aps,
floatfix,
]{revtex4-2}

\usepackage{graphicx}
\usepackage{dcolumn}
\usepackage{bm}
\usepackage{hyperref}
\usepackage{tikz}
\usepackage{tkz-euclide}

\usepackage{xcolor}

\begin{document}

\preprint{APS/123-QED}

\title{Stochastic Lorenz dynamics and wind reversals in Rayleigh-Bénard Convection}

\author{Yanni Bills}
\email{yanni.bills@yale.edu}
\affiliation{Department of Earth \& Planetary Sciences\\ Yale University, New Haven, Connecticut 06520, USA}

\author{John S. Wettlaufer}
 \email{john.wettlaufer@yale.edu, jw@fysik.su.se}
\affiliation{
Departments of Applied Mathematics, Earth \& Planetary Sciences, and Physics\\ Yale University, New Haven, Connecticut 06520, USA\\
 and Nordita, Royal Institute of Technology and Stockholm University, Hannes Alfvéns väg 12, SE-106 91 Stockholm, Sweden
}%


\date{\today}

\begin{abstract}
      \noindent The Lorenz equations \cite{Lorenz1963} are a severe Galerkin-truncation of the Oberbeck-Boussinesq (OB) equations describing Rayleigh-Bénard convection (RBC). Here we examine the mathematical connections between the chaotic lobe-switching behavior of a stochastic form of the Lorenz equations, that model the interaction between the thermal boundary layers and the core circulation, and the mean wind reversals in the experiments of Sreenivasan et al. \cite{Sreenivasan2002}.  Long-time numerical simulations of these stochastic equations, not easily accessible with the OB equations, yield a probability distribution for lobe inter-switch timings that exhibits non-Gaussian, multifractal behavior.  In the Gaussian frequency range the simulations mirror the laboratory measurements and the classical Hurst exponent and quadratic variation show Brownian second-moment statistics.  Further scrutiny reveals
      a non-linear cumulant generating function, or moment-exponent function, and thus multifractality.
       A simple generalized two-scale Cantor-cascade analysis reproduces these properties, showing that multiplicative intermittency, characteristic of turbulence, strongly influences the statistics. This demonstrates that this stochastic Lorenz system is a faithful, low-dimensional surrogate for mean-wind reversals in RBC.
\end{abstract}

\maketitle



\section{\label{sec:level1}Introduction}

The ubiquity and importance of thermal convection in technical and natural settings, particularly in the astro- and geophysical sciences, is very well known (e.g., \cite{Siggia1994, Manneville2006a, Bercovici2007, Doering2020}). The primary aim of studies of turbulent Rayleigh-B\'enard convection between two horizontal surfaces has been to determine the Nusselt number, $\operatorname{Nu}$, defined as the ratio of total heat flux to conductive heat flux, as a function of the Rayleigh number $\mathrm{Ra}$, the Prandtl number $\operatorname{Pr}$, and the aspect ratio of the convection cell $\Gamma$.  The principal control parameter is the Rayleigh number $\mathrm{Ra}=g \alpha \Delta T H^3 / \nu \kappa$, which is the ratio of buoyancy to viscous forces, where $g$ is the acceleration due to gravity, $\alpha$ the thermal expansion coefficient of the fluid, $\nu$ the kinematic viscosity, $\kappa$ the thermal diffusivity, and $\Delta T$ is the temperature difference across the fluid layer of depth $H$.  The Prandtl number, $\operatorname{Pr}=\nu / \kappa$, is considered a property of the fluid and the aspect ratio of the cell, $\Gamma$, defined as the ratio of its width to height, is considered a property of the experiment. For $\mathrm{Ra} \gg 1$ and fixed $\operatorname{Pr}$ and $\Gamma$, a relation is usually sought in the form of a power law: $\operatorname{Nu} = A(\operatorname{Pr}, \Gamma) \mathrm{Ra}^{\gamma}$.

\begin{figure}[h!]
    \centering
    \hspace*{-7.5mm}
    \includegraphics[width=1.0\linewidth]{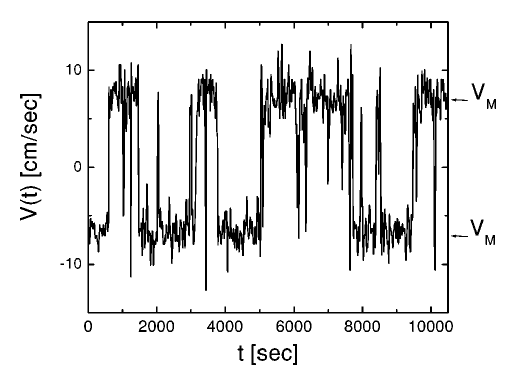}
    \caption{For $\mathrm{Ra}=1.5 \times 10^{11}$, the wind velocity as a function of time (for a total of about 3 hours) reproduced from Fig. 1 of Sreenivasan et al. \cite{Sreenivasan2002}.  They measured this by correlating two neighboring temperature probes on the center plane of the convection cell outside of the sidewall boundary layers.  They highlighted that the wind switches direction abruptly with no apparent order. The arrows on the right indicate the magnitude of the wind during its persistence in one direction. See \cite{Sreenivasan2002} for more details of the experiment.}
    \label{fig:a}
\end{figure}

Experiments, simulations, theory and analysis have focused on the nature of the power law at higher and higher $\mathrm{Ra}$.  Central here is how the boundary layers (BLs) at the cold upper and warm lower surfaces of the cell interact with the core flow in the interior; for a succinct history and description see Doering \cite{Doering2020} and references therein. An important component of this interaction is the ``mean wind'', which is the large-scale-circulation of the flow superimposed on the background turbulence.  The mean wind has a scale of order the experimental cell, and it is the transitions in the mean wind measured in a cylindrical cell by Sreenivasan et al. \cite{Sreenivasan2002} that we focus upon here.  Whether these transitions are wholesale reversals, that is flow cessation and reversal, or azimuthal rotations, we refer to them as reversals in the sense of the sign of the vertical velocity (see Figure \ref{fig:a} and the end of \S \ref{sec:level1}).

It is important to appreciate the nature of these experiments.  Prior to measuring the time series of reversals, Sreenivasan et al. \cite{Sreenivasan2002} held their apparatus at constant experimental conditions for nearly a month.  They then measured abrupt reversals of the mean wind as shown in their Figure 1 (reproduced here as Figure \ref{fig:a}), where they plot an increment of their time series of the mean wind for $\mathrm{Ra}=1.5 \times 10^{11}$. Finally, to generate reliable statistics, they recorded reversals continuously for 5.5 days.  Experiments were performed for $\mathrm{Ra}$ 
between about $10^8$ and $10^{13}$, but the most extensive for $\mathrm{Ra}=1.5 \times 10^{11}$, where they focused their analysis.

The velocity of the mean wind was about ${V_M} = 7 \mathrm{~cm} / \mathrm{s}$, the circumference of the apparatus was about $200 \mathrm{~cm}$, so on average the fluid traversed the cell in 30 s, which is long relative to the transition between the two mean wind directions seen in Figure \ref{fig:a}, which spans about 330 traversal times with the entire time series spanning some 16,000 traversal times.  In contrast, high resolution direct numerical simulations of the full Oberbeck-Boussinesq equations are run for 100-1000 turnover times and, depending on resolution and dimensionality, can take 10-20 days using parallel HPC methods on GPUs (See e.g., \cite{Tiwari2025}).  In this context the numerical challenge was addressed by Benzi and Verzicco \cite{Benzi2008} who effectively decreased $\operatorname{Pr}$ by artificially increasing thermal fluctuations, thereby allowing them to generate statistics for a more modest value of $\mathrm{Ra}=6 \times 10^5$.  They interpreted the transitions in terms of a bistable stochastic system, as did Sreenivasan et al. \cite{Sreenivasan2002}.  Another approach, using so-called augmented Lorenz equations, treats the random transitions in the rotation of a gas turbine \cite{waterwheel}, akin to the Malkus-Howard chaotic waterwheel, as an analogue to the wind reversals in the experiments of Sreenivasan et al. \cite{Sreenivasan2002}. 
Here we study the mathematical aspects of these wind reversal experiments using a stochastic variant of the canonical reduced model for convection due to Lorenz, which is derived directly from the Oberbeck-Boussinesq equations \cite{Lorenz1963, Saltzman1962}.

Experiments with a larger array of probes have characterized changes in the vertical velocity as either due to wholesale reversal (i.e., cessation and reversal) or azimuthal rotation \cite{Brown}.  However, Brown and Ahlers \cite{Brown} define an average rate of occurrence of reorientations, and note that it (a) depended strongly on the definition of the reorientation parameters, and (b) was extremely sensitive to minor changes of the apparatus.  Therefore, while we recognize the existence of reversals and rotations, we remain agnostic concerning their relative occurrence in the experiments of Sreenivasan et al. \cite{Sreenivasan2002} and, as noted above, we refer to the transitions in the vertical velocity in Figure \ref{fig:a} as reversals.

\section{\label{sec:level2}A Stochastic Surrogate of Experiments}

The deterministic Lorenz 1963 model captures many key features of RBC in a symmetry-invariant, three-mode representation of the full Oberbeck-Boussinesq equations \cite{Lorenz1963}. Importantly for the phenomenology of the BL-core flow interaction, the $Z$ variable represents the deviation of the vertical temperature profile from linearity, and arises perturbatively as a closure for the non-linear Jacobian in the temperature equation. 
 Deterministically, $Z$ balances the diffusive relaxation of the BL with non-linear forcing by convection. When positive, the vertical temperature profile is steeper near the BLs and shallower in the bulk. In an experiment or full DNS of the OB equations, the BLs are continuously perturbed by the large-scale circulation, the small-scale turbulence and plume detachments and instabilities. Such processes trigger multi-scale fluctuations in the vertical gradient field.  The introduction of Gaussian white noise (GWN) forcing in the $Z$-equation provides a direct means of modeling these BL-core flow interactions, which is the principal reason for the stochastic Lorenz equations we study in Eqs. \eqref{eq:SL}.

The model we examine in a different regime than Coti Zelati and Hairer \cite{Zelati2021} is the deterministic Lorenz system with additive GWN in the third component as follows,
\begin{equation}
    \begin{aligned}
        &\frac{dX}{dt}=\sigma (Y-X),\\
        &\frac{dY}{dt} = X(\rho-Z) -Y\qquad\textrm{and}\\
        &\frac{dZ}{dt} = -\beta Z+XY+\hat{\alpha}\xi,
    \end{aligned}
    \label{eq:SL}
\end{equation}
where, $X$ represents the intensity and direction of the convective roll, $Y$ ($Z$) represents the deviation of the horizontal (vertical) temperature field from the pure conductive state with a linear temperature profile. The Prandtl number is $\sigma$ taken to be 10, $\beta$ is a geometric parameter equal to 8/3, and $\rho$ is the reduced Rayleigh number that characterizes the nature of the flow regime, heuristically like $\mathrm{Ra}$ for the full OB equations.  The amplitude of the additive noise is $\hat{\alpha}$, and $\xi$ is an uncorrelated, stationary, Gaussian process that is treated as a sequence of Brownian increments.

When $\hat{\alpha}=0$ and $\rho<1$ ($\rho>1$) there is a unique global attractor at the origin (there are two non-trivial fixed points, which become unstable when $\rho\approx24.74\equiv\rho_c$, either exhibiting a chaotic attractor or stable limit cycles).
Coti Zelati and Hairer \cite{Zelati2021} investigated a bifurcation threshold for the system ($\rho<1$) as $\hat{\alpha}$ increased. 
Thus, there is a transition from a single unique ergodic invariant measure to two ergodic invariant measures with increasing noise strength.
{\color{black} We study the regime with $1<\rho<\rho_c$. }

{\color{black} The noise-induced transition discovered by Coti Zelati and Hairer \cite{Zelati2021} is analogous to the measure-non-uniqueness phase transition described in \cite{Hairer2006} and \cite{Stacy2026}. At the level of the Galerkin truncations, thus including the stochastic Lorenz 1963 (Eqs. \ref{eq:SL}) and 1996 models, the phase transition behavior captured by these lower-dimensional stochastic differential equations (SDEs) is conjectured to extend to the full two-dimensional stochastic Navier-Stokes (NS) equations \cite{Stacy2026}. Moreover, we note that Eqs. \eqref{eq:SL} act as a model of Gaussian-forced NS flow on a torus \cite{Zelati2021}. This argument is furnished by analysis of a three-dimensional system of SDEs (very similar to Eqs. \ref{eq:SL}) in \cite{Brz2022}, whereby these equations and the stochastic two-dimensional NS equations can be rewritten in the same form (see Remark 2.1 and Eq. 2.1 in \cite{Brz2022}). This similarity, and subsequent connection to the stochastic two-dimensional NS equations, is made concrete in Remark 1.5 of \cite{Zelati2021}; the transverse-instability route to non-uniqueness of invariant measures is present in both Eqs. \ref{eq:SL} and the system of SDEs studied in \cite{Brz2022}. Thus, the mathematical link between Eqs. \ref{eq:SL} and the stochastic two-dimensional NS equations bolsters the physical mechanisms discussed in the beginning of this section.}


\subsection{Transformed Stochastic System}

We use Coti Zelati and Hairer's \cite{Zelati2021} transformation of Eqs. \eqref{eq:SL} viz.,
\begin{equation}
    \begin{aligned}
        &\chi=\frac{2}{1+\sigma}, \quad \eta= \frac{1+\sigma}{2\sigma}, \quad \gamma=\chi \beta, \quad \nu^2 = \chi^5 \sigma, \\
        &\alpha =\nu \sqrt{\sigma} \hat{\alpha}, \quad \text{and} \quad z_\star=2+\chi^2 \sigma(\rho -1).
    \end{aligned}
\end{equation}
This allows for the change of variables,
\begin{equation}
    \begin{aligned}
        &x(t)=\frac{\nu}{\chi}X(\chi t), \quad y(t)=\nu \sigma(Y(\chi t)-X(\chi t)), \\
     \textrm{and~}&z(t)=z_\star - \chi^2 \sigma Z(\chi t),
    \end{aligned}
\end{equation}
which transforms Eqs. \eqref{eq:SL} to
\begin{equation}
    \begin{aligned}
        &\dot{x}=y, \\
        &\dot{y} = x(z-2)-2y, \qquad\textrm{and}\\
        &\dot{z}=-\gamma (z-z_{\star})+\alpha \xi - x(x+\eta y).
        \label{eq:SLT}
    \end{aligned}
\end{equation}

Here we make two comments about the properties of Eqs. \eqref{eq:SLT}.
Firstly, according to Theorem 3.1 of \cite{Zelati2021}, for all parameter values, Eqs. \eqref{eq:SLT} admit at least one, and at most two, ergodic invariant probability measure(s).  For $\rho$ and $\hat{\alpha}$ sufficiently large, the system admits a non-trivial measure and thus long time averages of individual trajectories will eventually reflect the ensemble average.
Secondly, when linearized, the evolution of $z$ in Eqs. \eqref{eq:SLT} is an Ornstein-Uhlenbeck (OU) process and thus governed by a Gaussian stationary probability density. Therefore, to remain in the OU regime we choose parameter values such that $|x(x+\eta y)|$ remains small.

\subsection{Experimental Surrogate}

Our \textit{ansatz} is to ascribe the sign changes in $X$, or ``lobe switching'' on the Lorenz attractor, to the experimental reversals in the mean-wind shown in Fig.~\ref{fig:a}. Therefore, due to the essential role of the BL-core flow interaction discussed above, we examine the behavior of Eqs. \eqref{eq:SLT}.

\begin{figure}
    \centering
    \hspace*{-7.5mm}
    \includegraphics[width=1.0\linewidth]{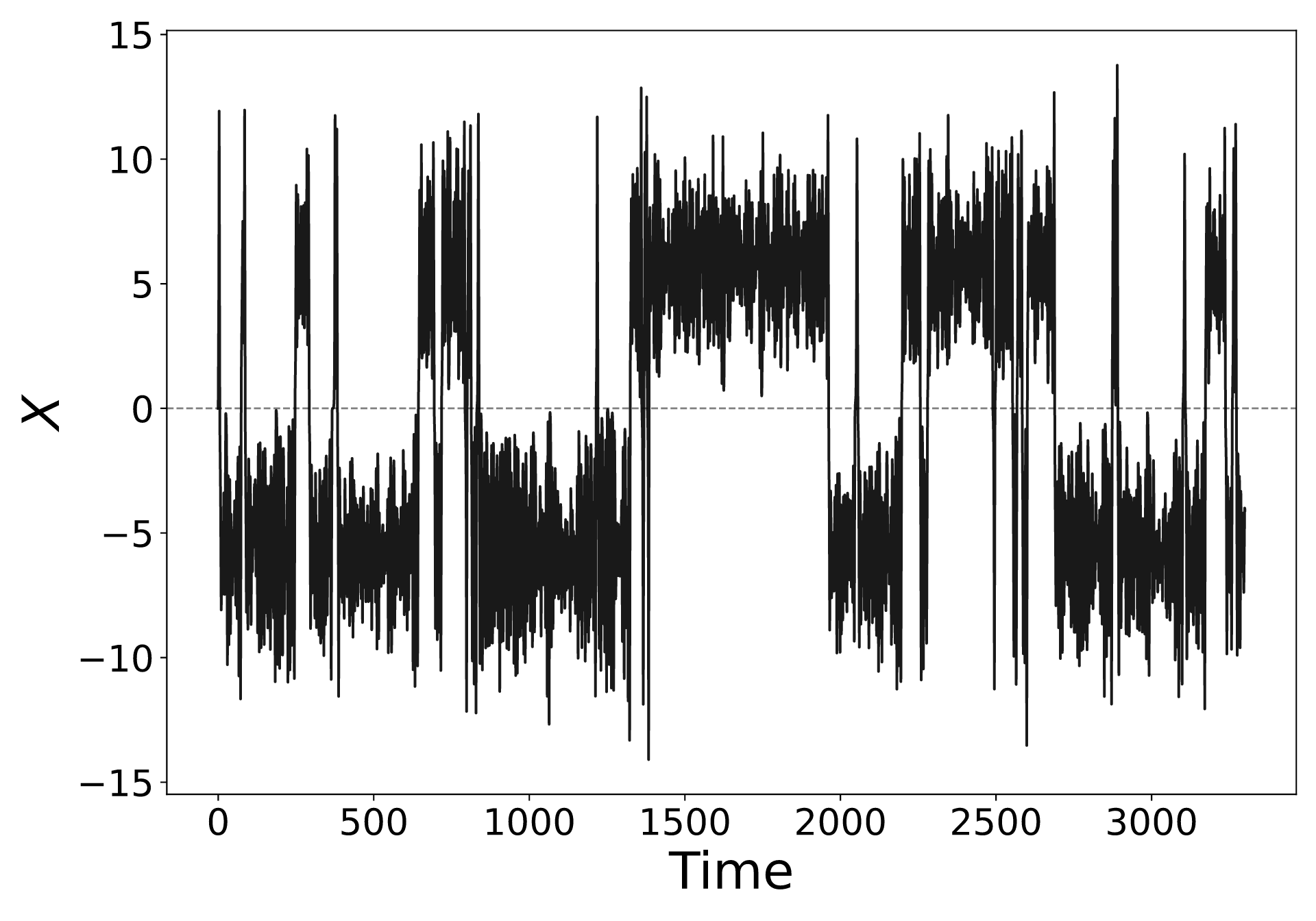}
    \caption{The time series of $X$ after being back transformed from a realization of Eqs. \eqref{eq:SLT} by a non-adaptive Euler-Maruyama scheme. The simulation time is $600/\chi$, the reduced Rayleigh number is $\rho=14$, and the noise amplitude is $\hat{\alpha}=5$. Compare with Fig. \ref{fig:a}.}
    \label{fig:1}
\end{figure}

The experimental time series is reminiscent of the early chaotic regime for the deterministic $X(t)$, where the dwell time is off of $X=0$. However, for additive noise in all three components, there is a non-monotonic coupling between noise and unstable periodic orbits for $\rho \approx \rho_c\approx 24.74\,$, {\color{black}observed for $\hat{\alpha}\in \{1,2,\dots,9,10\}$ \cite{Agarwal2016}.} Thus,  for $\rho=14$ and {\color{black}a moderate noise amplitude of} $\hat{\alpha}=5$, we examine the time series of $X(t)$ from Eqs. \eqref{eq:SLT} as shown in Fig. \ref{fig:1}. {\color{black} At $\rho=14$ in the deterministic Lorenz system, trajectories evolve into one of two symmetric fixed points, meaning that persistent reversals do not occur. It is the advent of stochasticity in the $Z$ equation that activates a chaotic invariant set of initial conditions (i.e., with positive Lebesgue measure) in the dynamical phase space, thereby creating flow reversals at $\rho=14$. When $\rho$ ($\hat{\alpha}$) is fixed, increasing $\hat{\alpha}$ ($\rho)$ decreases the mean time between successive reversals. This behavior exists in concert with the fact that in RBC, for a given Pr, chaotic and/or periodic reversals only occur when Ra is sufficiently large (see Fig. 3 in \cite{Araujo}).} {\color{black} The parameters used for our investigation remain fixed at the aforementioned values of $\rho=14$, $\hat{\alpha}=5$, $\sigma$ = 10 and $\beta$ = 8/3.}

\subsection{Timing of the Reversals}

We follow Sreenivasan et al. \cite{Sreenivasan2002} who defined a discrete process, $T_n$, as the time between zero and the $n^{\text{th}}$ lobe switch. We then consider a stochastic process,
\begin{equation}
G_n:=T_n-(an+b),
\label{eq:Gn}
\end{equation}
where $a$ and $b$ define the best fit line for $T_n$ versus $n$. Thus, $G_n$ is the deviation from linearity of the relationship between $T_n$ and $n$, and we compare this with the experiment.

\subsection{Numerical Simulations}

\begin{figure}
    \centering
    \hspace*{-5mm}
    \includegraphics[width=1.00\linewidth]{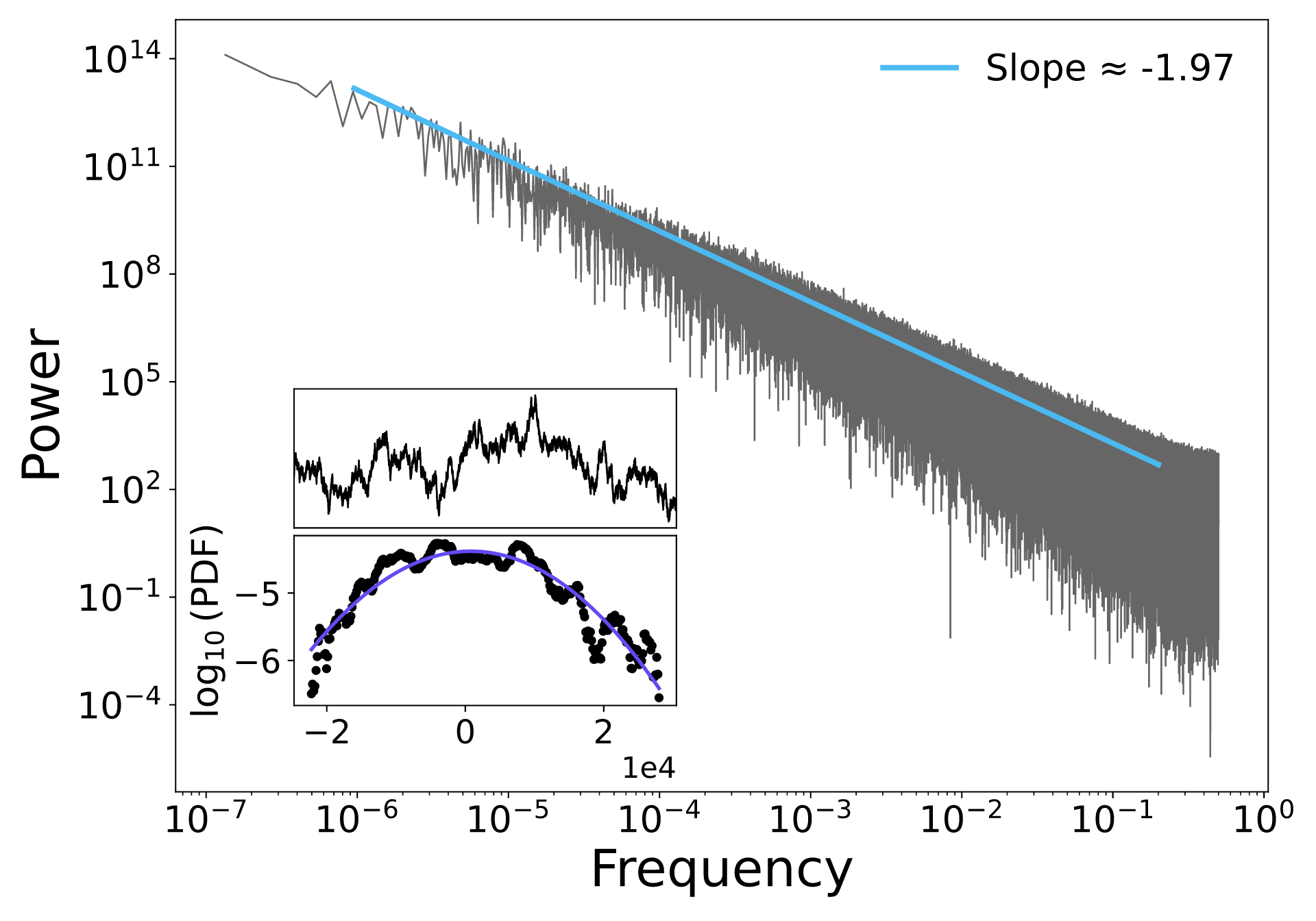}
    \caption{The PSD versus the frequencies (cycles per switch) of the process $G_n$ with the light blue line {\color{black}giving the best fit linear regression} as a visual for the decay. The bottom-left inset displays the raw $\log_{10}$(PDF) of the same process, with the indigo curve being the expected quadratic fit for a true Gaussian. The black line graph above the PDF is the full time series data of $G_n$.}
    \label{fig:3}
\end{figure}

We perform long-time numerical simulations of Eqs. \eqref{eq:SLT} for $\rho=14$ and $\hat{\alpha}=5$.  A run time of $T= 120000000/\chi$ generates 7.4 million lobe switches.  The raw $\{G_n\}_{n=1}^N$ data of Sreenivasan et al. \cite{Sreenivasan2002} show a Gaussian probability density function (PDF), with a clear quadratic fit and a power spectral density (PSD) exhibiting a decay of approximately -2.  
 In Fig. \ref{fig:3} we show the simulated PSD with a decay of $-1.97$ across multiple decades---a signature of Gaussianity. However, the inset shows that the raw $\log_{10}$(PDF) is not Gaussian. Given the statistical size of our simulations, the fluctuations in the PDF are real; increasing the number of bins only emphasizes their departure from normality.

\subsection{Frequency Filtering}

\begin{figure}
    \centering
    \hspace*{-4mm}
    \includegraphics[width=1.0\linewidth]{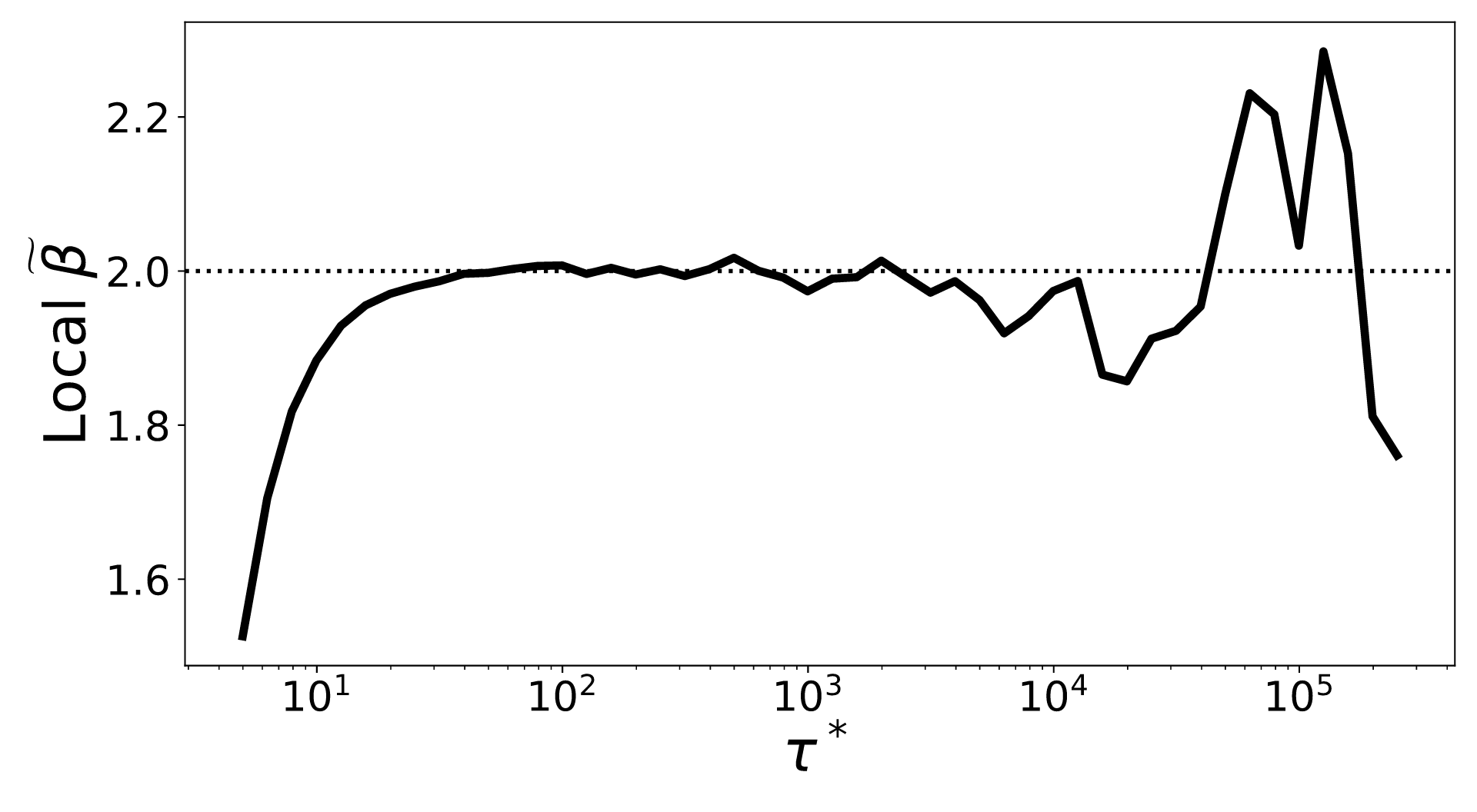}
    \caption{The power law exponent, $\widetilde\beta$, across a wide range of time-scales (inverse frequencies). Window sizes for exponent approximation are 0.75 decades wide, with step sizes between tiles being 0.1 decades wide.}
    \label{fig:5}
\end{figure}

\begin{figure}
    \centering
    \hspace*{-6mm}
    \includegraphics[width=1.0\linewidth]{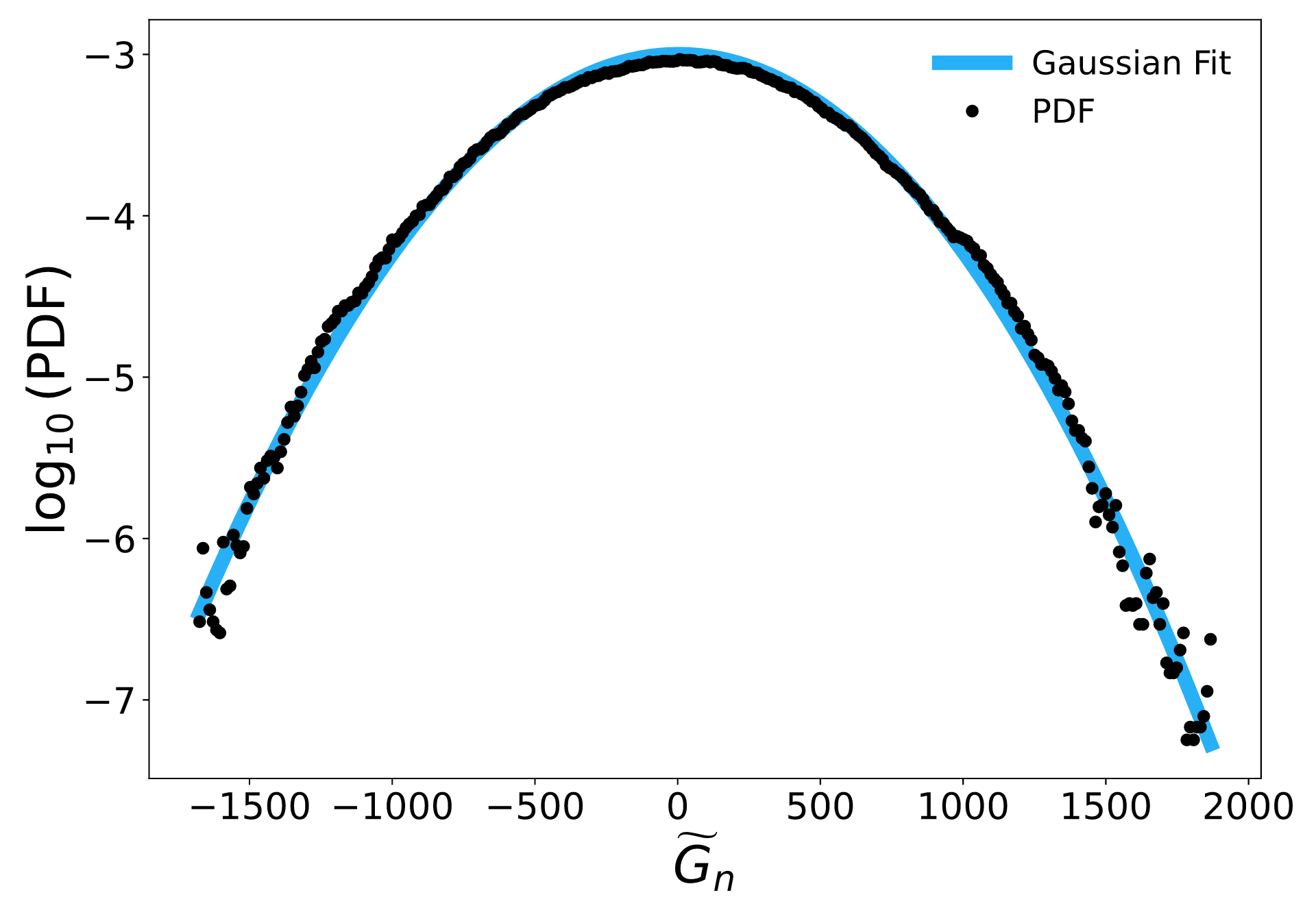}
    \caption{The $\log_{10}$(PDF) of $\widetilde{G}_n$ under the same parameter regime---only the values of $G_n$ within the $\left[10^{-4},10^{-2} \right]$ cycles per switch frequency range (in frequency space) are kept. }
    \label{fig:6}
\end{figure}

First, we use Fig. \ref{fig:3} to compute the band-limited power spectral density; $\text{PSD}_{\text{band-limited}}\propto f^{-\widetilde\beta}$, such that $\widetilde\beta\approx 2$. Using a window width of 0.75 decades we tile the full PSD to find that the time-scale range of $[10^2,10^4]$ (equivalent to the [$10^{-4},10^{-2}$] frequency range through $f=1/\tau^*$) satisfies $\widetilde\beta\approx 2$,  as seen in Fig. \ref{fig:5}. Second, we employ an inverse Fourier transform with an appropriate band-pass filter for this frequency range. The filtered $\widetilde{G}_n$ sequence is then generated with the PDF shown in Fig. \ref{fig:6}, which clearly shows Gaussianity.

This frequency filtering is our method of replicating the sampling frequency in the experiments of Sreenivasan et al. \cite{Sreenivasan2002}.  We view their mean-wind reversal measurements as macroscopic changes of state with high-frequency background fluctuations reflected in the raw data of $G_n$, but filtered by the probe sampling rate, as reflected in the characteristic time of 30 seconds for a non-spurious mean-wind reversal transect of the apparatus. Experimentally, there is a physical and measurement lower limit on reversal transect time, and there is a low-frequency limit associated with the total of about 2,300 measured mean-wind reversals. In contrast, due to the extremely long record of switches, the lobe inter-switch timings produced by the stochastic Lorenz simulations, encompass a far wider range of frequencies (cycles/switch), which produce the non-Gaussian structure of the raw PDF in the inset of Fig. \ref{fig:3}. Therefore, the experimentally observed Gaussianity may be due to the intrinsic coarse-graining of the measurements, which do not fully resolve the turbulence, suggesting that the frequency range producing Gaussianity in the simulations is a scale dependent property of the convection. 

\section{Simulation of Second Moment Statistics}

\subsection{Hurst Exponent}

\begin{figure}
    \centering
    \hspace*{-5mm}
    \includegraphics[width=1.0\linewidth]{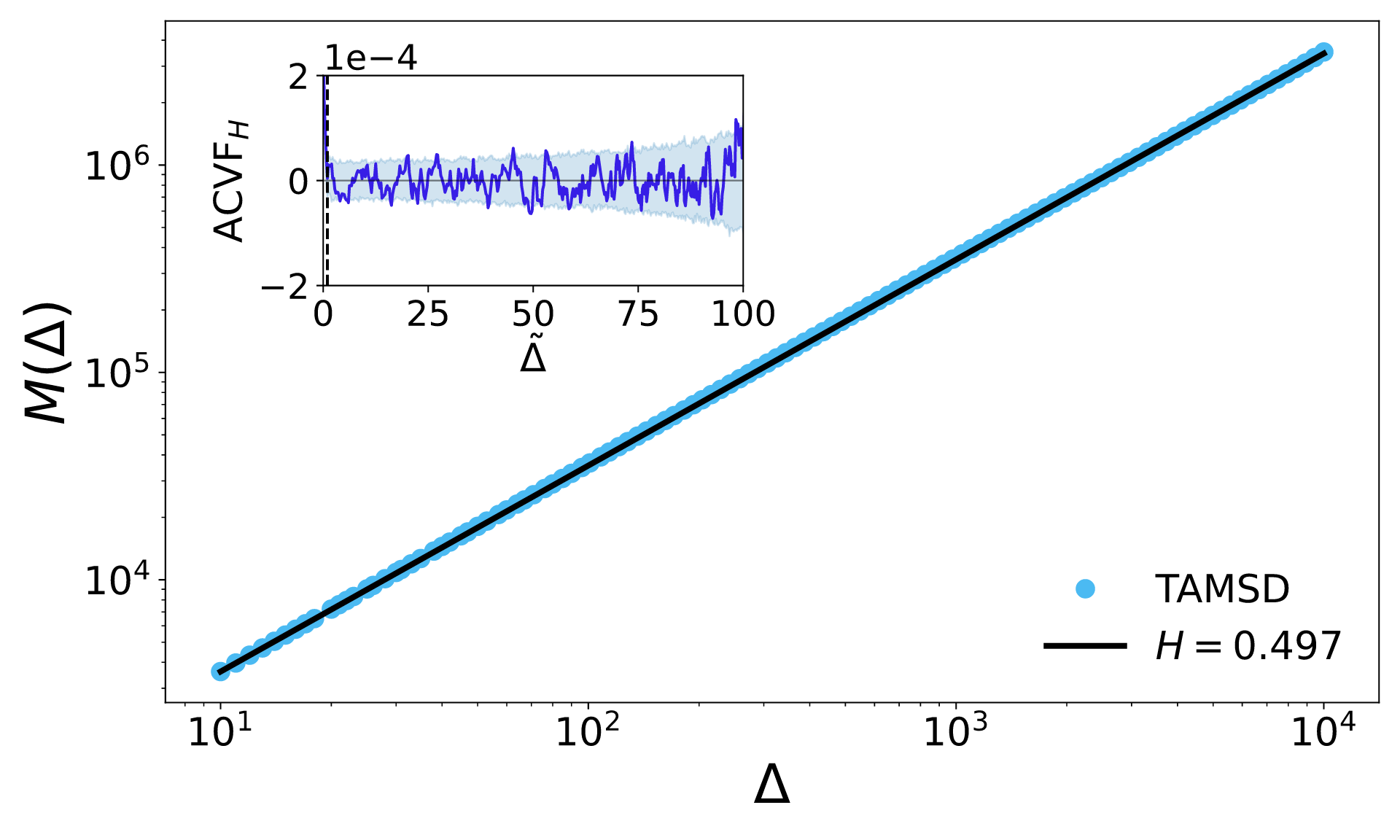}
    \caption{A log-log plot of TAMSD versus lag time $\Delta$, the slope of which is 2$H$, where $H$ is the classical Hurst exponent. The inset is the ACVF of local $H$ estimates versus a rescaled lag time $\tilde{\Delta}$ with the pale-blue band giving the 95\% confidence interval.}
    \label{fig:12}
\end{figure}

{\color{black} For a lag time $\Delta\in \mathbb{R}_{\geq 0}$, the time averaged mean square displacement (TAMSD) is defined as \cite{Balcerek} }

\begin{equation}
    M(\Delta)=\frac{1}{N-\Delta}\sum_{n=1}^{N-\Delta}(G_{n+\Delta}-G_n)^2,
\end{equation}
where in the case of fractional Brownian motion (FBM), $M(\Delta)\propto \Delta^{2H}$, with $H$ the classical Hurst exponent.  
Thus, $H$ is half the slope of the TAMSD vs. $\Delta$ on a log-log plot. As shown in Fig. \ref{fig:12}, the slope $2H\approx0.994$ giving a classical Hurst exponent of $0.497$.

To verify the FBM properties of $G_n$, we compute local $H$ estimates for overlapping windows of length $L=2^{16}$. Within each window, the TAMSD and $M(\Delta)$ are populated such that local $H$ values are determined from the $\Delta^{2H}$ scaling law. The $H$ of each window appends the list, $H_i$, of length $N_w$, which is the number of windows. The autocovariance function (ACVF) for these local $H$ estimates is given by
\begin{equation}
    \text{ACVF}_H(k)=\frac{1}{N_w-k} \sum_{i=0}^{N_w-k-1}(H_i-\bar{H})(H_{i+k}-\bar{H}),
\end{equation}
where $\bar{H}$ is the average of all the $H_i$. We then plot the ACVF versus a rescaled lag time, $\tilde{\Delta}=k/4$, in the inset of Fig. \ref{fig:12}, showing oscillations about zero, which is a characteristic of FBM \cite{Balcerek}. Thus, the $H$ estimate does not drift, thereby allowing us to use the power law for a global estimate of the classical Hurst exponent. {\color{black} This is further supported as we find that $\bar{H}\approx 0.5$ as well, meaning that within each length $L$ window, local Hurst exponents are consistently near $1/2$.

}

\subsection{Increment Testing}

A classical Hurst exponent of 1/2 defines a Wiener process to second moment.  Namely, the process defines white noise with uncorrelated increments and no persistence. A Wiener process is also referred to as standard Brownian motion. {\color{black}An important takeaway of this is that, to the second moment, $G_n$ is Markovian (memoryless), which implies that the flow reversals themselves can take longer or shorter than the mean inter-switch time with probability independent of anything further in the past than the most recent lobe switch. 

If $H$ were less than 1/2, $G_n$ would be anti-persistent and thus the probability law governing reversals would reflect very short inter-switch timings followed by very long ones. This also eliminates Markovianity from the reversal process. Physically, such a scenario is consistent with the mean wind having a preference for one orientation over another. 

On the other hand, were $H>1/2$, $G_n$ would be persistent, and thus the probability law governing reversals would reflect that longer (shorter) than usual inter-switch times are more likely to be followed by yet another longer (shorter) than usual inter-switch time. This implies that the mean wind might admit a form of bistability in terms of inter-switch timings. Physically, this could correspond to a boundary layer growth and breakdown occuring on a time scale slower or of order the circulation time scale. 

In summary, the mean wind does not behave like either of the $H\neq 1/2$ cases above. We have verified that it has no preference for any particular orientation and does not clearly exhibit bistability of the inter-switch timings themselves. Clearly there is interpretable physical meaning associated with this behavior, and hence}  
now we test $G_n$ for further properties of a lattice-sampled Wiener process in the $L^2$ sense.

\subsubsection{Variance}

As described in Eq.~\eqref{eq:Gn}, $T_n$ is a discrete-time process with uniform time steps that are non-negative integers, and so $G_n$ is continuous by construction. The base increments of $T_n$ are of size $\Delta t_{\text{base}}=1$.  
To determine the finite difference variance, we impose

\begin{equation}
\Delta t_k:=k \Delta t_{\text{base}}=k.
\end{equation}
Let $k \in \{2^0,2^1,2^2,2^3,...,2^{10} \}$ and define
\begin{equation}
    \begin{aligned}
    v_n^{(k)}:=&\frac{G_{n+k}-G_n}{\Delta t_k} = \frac{G_{n+k}-G_n}{k},\\ \qquad n&=1,...,N-k,
    \end{aligned}
\end{equation}
where index $n$ marks the start of a measured window, all of which overlap with one another. For a fixed $k$ we compute the variance $\hat{\sigma}^2(k)$ as

\begin{equation}
    \hat{\sigma}^2(k) =\frac{1}{N-k-1} \sum_{n=1}^{N-k} \left(v_n^{(k)}-\bar{v}^{(k)} \right)^2,
\end{equation}
where $\bar{v}^{(k)}$ is the average of $v_n^{(k)}$ over all $n$ for a particular $k$. {\color{black} Note that $v_n^{(k)}$ is effectively a discretized derivative of $G_n$. Thus, the relevant variance quantity, $\hat{\sigma}^2$, is a variance of derivatives and can be interpreted as a variance of the nonlinear behavior of displacements for $G_n$.}

\begin{figure}
    \centering
    \hspace*{-4mm}
    \includegraphics[width=1.00\linewidth]{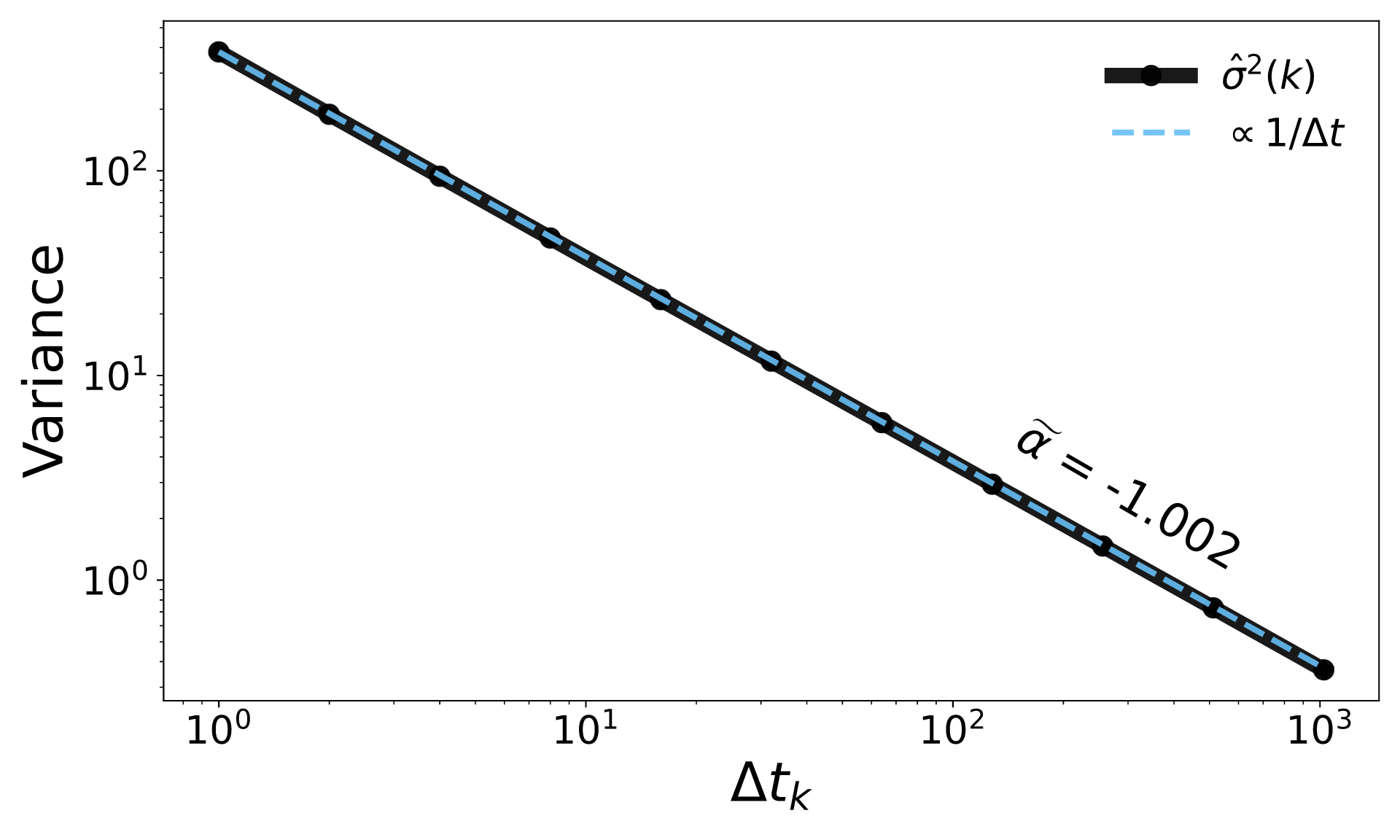}
    \caption{A log-log plot of the sample variance $\hat{\sigma}^2(k)$ versus $\Delta t_k=k$ for $k\in\{2^0,2^1,2^2,2^3,...,2^{10} \}$.}
    \label{fig:14}
\end{figure}

For a Wiener process, $W(t)$, we have 
\begin{equation}
    \begin{aligned}
        \text{Var} \left(v_n^{(k)} \right) &= \frac{\text{Var}(W(t_{n+k})-W(t_n))}{(\Delta t_k)^2}\\
        &= \frac{\Delta t_k}{(\Delta t_k)^2}=\frac{1}{k},
    \end{aligned}
\end{equation}
so we seek $\hat{\sigma}^2(k) \propto k^{-1}$, the veracity of which is determined by estimating $\widetilde\alpha$, where $\hat{\sigma}^2(k)\propto (\Delta t_k)^{\widetilde\alpha}$, and

\begin{equation}
        \widetilde\alpha = \frac{d \log_{10}(\hat{\sigma}^2(k))}{d\log_{10}(\Delta t_k)}.
\end{equation}
Fig. \ref{fig:14} shows our estimate $\widetilde\alpha=-1.002$ from $G_n$.  Further confirmation that we find a Wiener process is the divergence of the finite difference variance as $\Delta t_k \rightarrow0$.

\subsubsection{Quadratic Variation}

For standard Brownian motion with mean $\mu$ and variance $\sigma^2$, the quadratic variation (QV) is $\sigma^2 t$ with probability 1, a consequence of which is that $dW(t)^2=dt$. The process $\Delta_1 G_n:=G_{n+1}-G_n$, for $n=1,...,N-1$, has a one-step sample variance given by
\begin{equation}
    \hat{\sigma}^2=\frac{1}{N-1}\sum_{n=1}^{N-1}\left(\Delta_1 G_n-\overline{\Delta_1 G}\right)^2,
\end{equation}
where $\overline{\Delta_1 G}$ is the mean over $n$ of $\Delta_1 G_n$. We now fix a window length $L_{\text{samples}}=2^{20}$, where $k$ is sampled from $[1,2^{20}]$ such that each decade of $k$ has the same number of points. The mesh inside one of the equal length windows is $\Delta t_k=k$, with the number of observed increments being $m_k=\lfloor(L_{\text{samples}})/k\rfloor$. The record is tiled with non-overlapping windows that have starting indices $S_k=\{0,L_{\text{samples}},2L_{\text{samples}},...,q_{\max}L_{\text{samples}}\}$ where $q_{\max}=\lfloor \frac{N+1-L_{\text{samples}}}{L_{\text{samples}}} \rfloor$. Thus, for every $s \in S_k$, we build a $k$-spaced grid $t_j=s+jk$ for $j=0,...,m_k$ and form

\begin{equation}
    QV_k(s)=\sum_{j=0}^{m_k-1}\left( G_{t_j}-G_{t_j-jk} \right)^2\equiv\sum_{j=0}^{m_k-1}(\Delta_kG_{t_j})^2.
\end{equation}

\begin{figure}
    \centering
    \hspace*{-4mm}
    \includegraphics[width=1.0\linewidth]{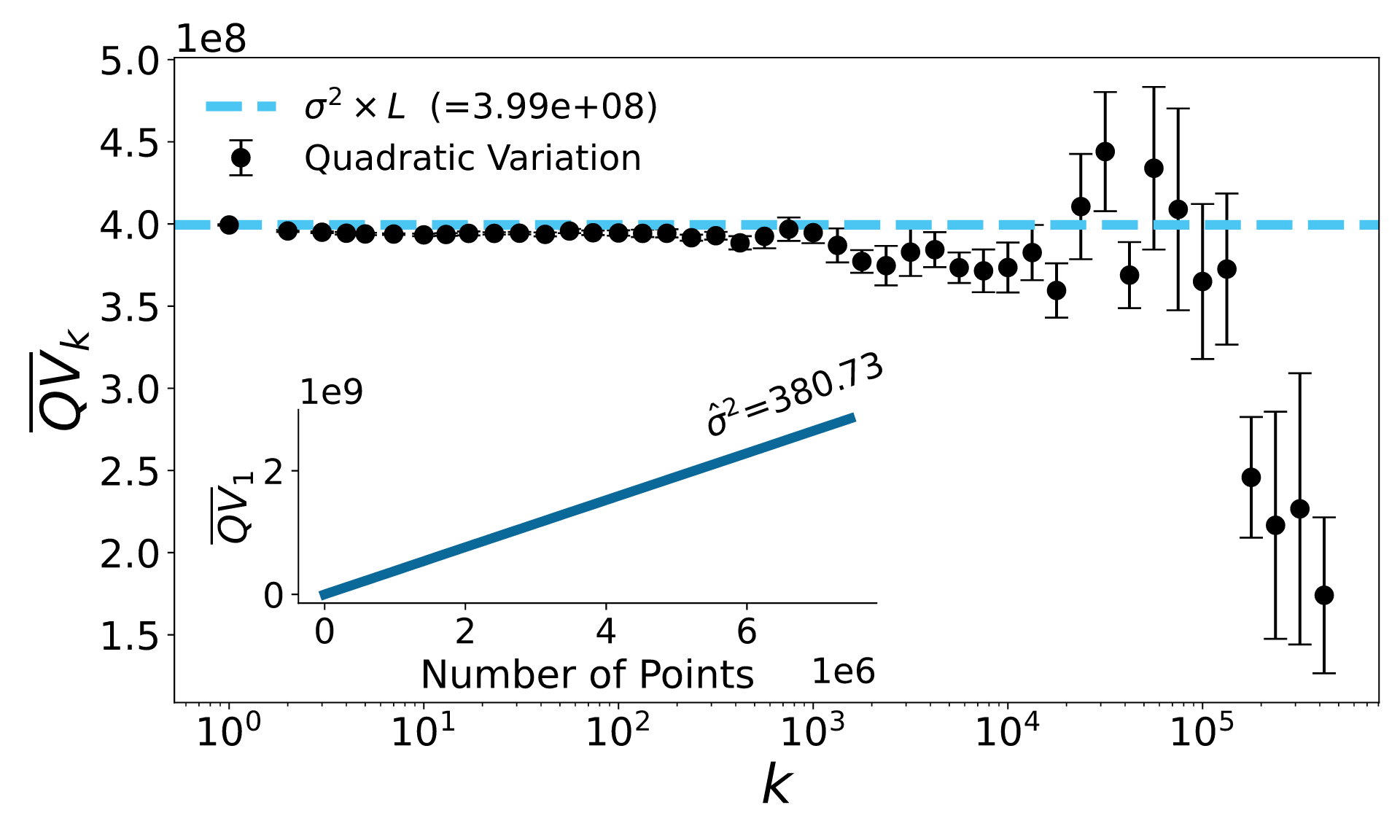}
    \caption{The mean quadratic variation plotted versus grid sizing parameter $k$.  The dashed, light blue line is the convergence target (estimated one-step variance), and the black dots are the computed values for each $k$, where the error bars are the standard error of the mean for each window for a particular value of $k$. The inset shows the linear relationship between the $k=1$ quadratic variation, $\overline{QV_1}$, and the points summed over in determining the full $\{G_n\}_{n=1}^N$.}
    \label{fig:15}
\end{figure}

In Fig. \ref{fig:15} we plot $\overline{QV}_k$, which is the empirical mean of $QV_k(s)$ over $s$. Clearly, as $k\rightarrow0$, 
$\overline{QV}_k$ converges to $\hat{\sigma}^2 L_{\text{samples}}$, here within $2.5\%$. As $k$ grows, the number of entries per window of size $L_{\text{samples}}$ decreases substantially, introducing the wider margins of error.  Moreover, the inset shows $\hat{\sigma}^2$-linear growth with the cumulative sum of points for the $k=1$ quadratic variation $\overline{QV_1}$. This shows the consistency of QV over the length of $G_n$.

\subsubsection{White Noise Increments}

\begin{figure}
    \centering
    \vspace*{0mm}
    \includegraphics[width=1.0\linewidth]{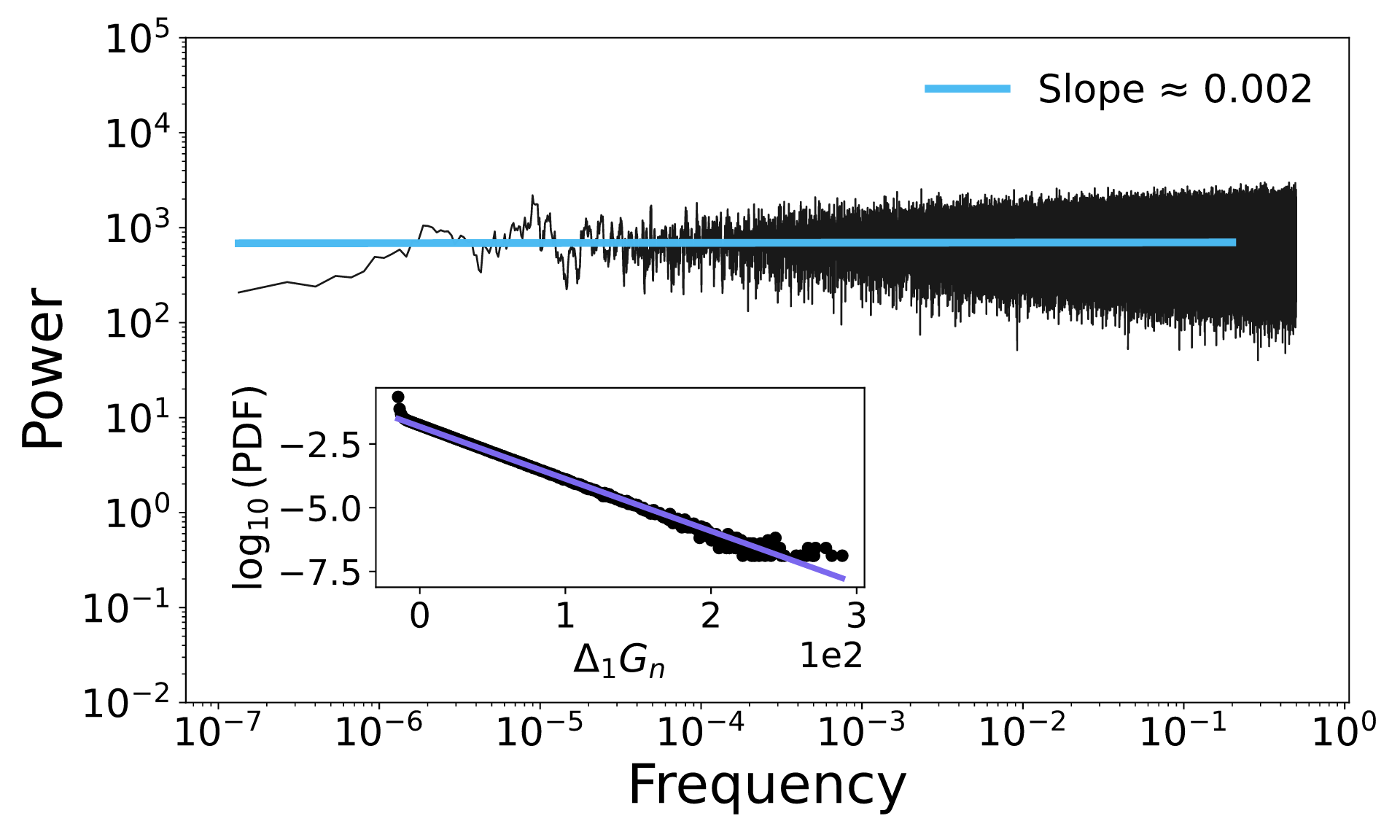}
    \caption{The PSD of $\Delta_1 G_n$ is plotted in black and the light blue line is a {\color{black}best fit linear regression producing the slope} $\sim0.002$. The inset shows the $\log_{10}$(PDF) of the increment process (black dots) with a linear fit (indigo line) implying an exponential distribution.}
    \label{fig:16}
\end{figure}

Finally, we show that Eqs.~\eqref{eq:SLT} produce white noise increments. In Fig. \ref{fig:16} we plot the PSD of $\Delta_1 G_n$ for the entire simulation dataset. Firstly, the spectrum is flat over nearly seven decades showing that the increments of $G_n$ are uncorrelated.  Secondly, the process $\Delta_1 G_n$ has an approximately zero mean, it is invariant under arbitrary time shifts, and its variance and autocovariance do not drift. These properties show that the increments are wide-sense stationary. Thirdly, a consequence of these properties and the Wiener-Kinchin theorem suggest that the increments are indeed uncorrelated through the relation between the PSD and the autocovariance. The corpus of these properties show that $G_n$ is a lattice-sampled Wiener process in the $L^2$ sense.

However, these increments are not normally distributed--the PDF of the increments shown in the inset of Fig. \ref{fig:16} are best fit by an exponential distribution. It is notable that whereas Sreenivasan et al. \cite{Sreenivasan2002} measured Brownian behavior for $T_n$, they also found an exponential distribution for the reversal time increments (see Fig. 5 in \cite{Sreenivasan2002}) and the return times.

\section{Multifractal Behavior}

\subsection{Numerical Comparison with Experiment}

\begin{figure}
    \centering
    \includegraphics[width=1.0\linewidth]{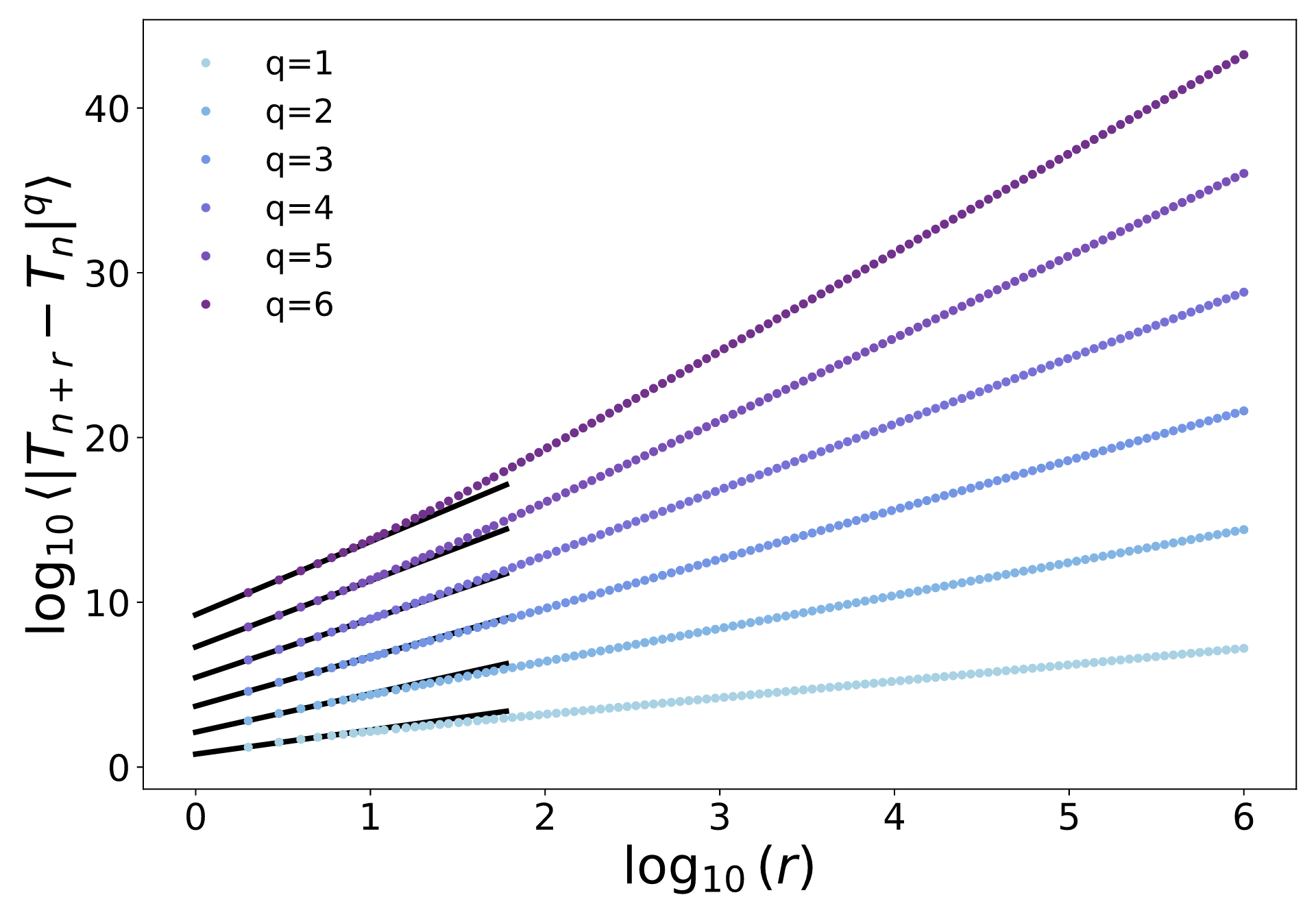}
    \caption{The $q$-moment generalized inter-switch intervals on log-log axes are plotted against lag times, $r$. Each moment from $q=1$ to $q=6$ is shown in the blue to purple gradient. The black lines indicate the initial, short-lag-time scaling laws. Cf. Fig. 8 of \cite{Sreenivasan2002}.}
    \label{fig:Tnr}
\end{figure}

\begin{figure}
    \centering
    \includegraphics[width=1.0\linewidth]{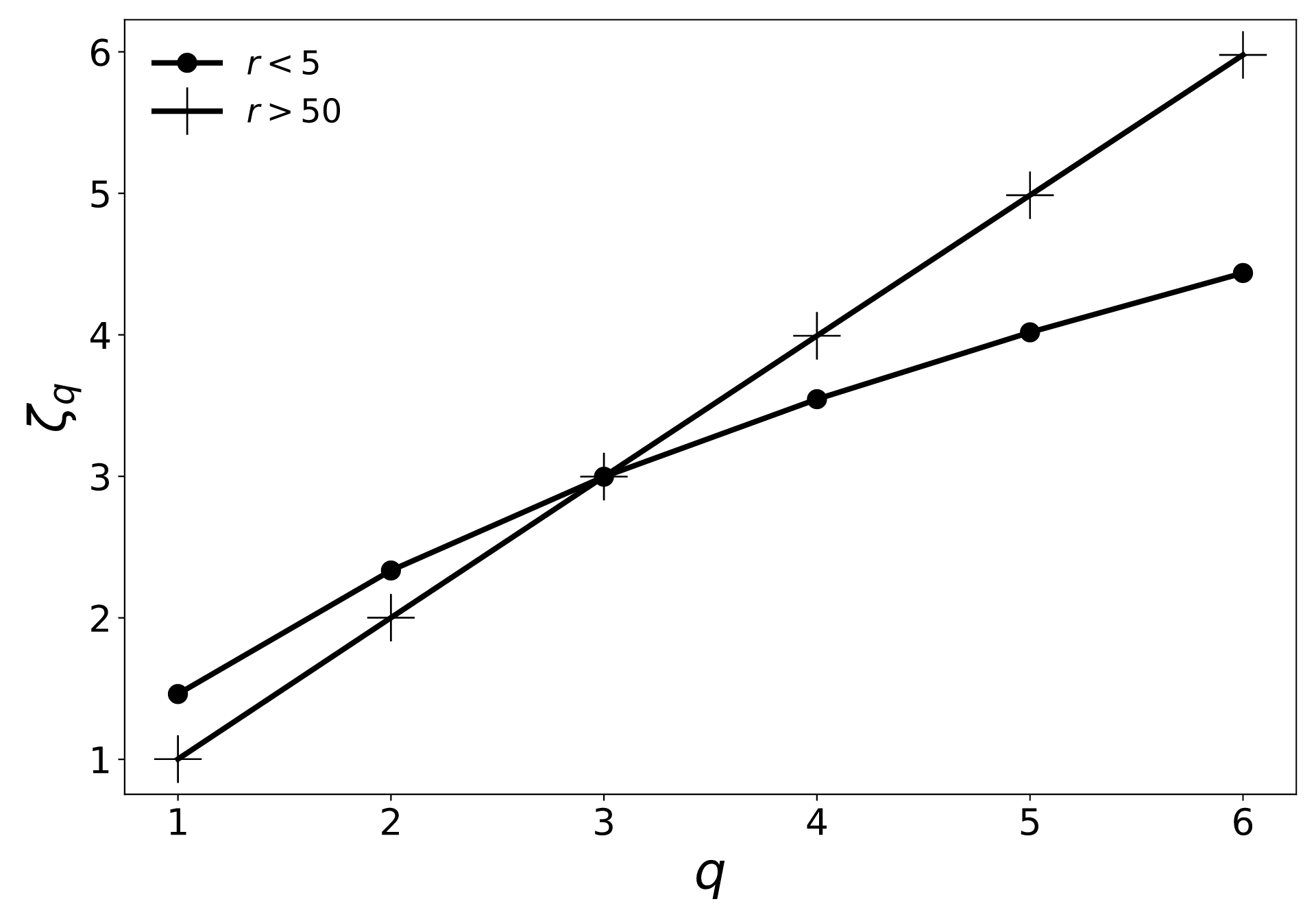}
    \caption{The local power law exponents, $\zeta_q$, are plotted against $q$ for the $r<5$ regime (circles) and $r> 50$ regime (crosses).  Cf. Fig. 9 of \cite{Sreenivasan2002}.}
    \label{fig:ZetaQ}
\end{figure}

Sreenivasan et al. \cite{Sreenivasan2002} defined the generalized inter-switch intervals as
\begin{equation}
    \Delta T_r=T_{n+r}-T_n, 
    \label{eq:Gswitch}
\end{equation}
and in Fig. \ref{fig:Tnr} we show the relationship between $\log_{10}\langle |\Delta T_r |^q \rangle$ and $\log_{10}r$ for $q\in\{1,2,...,6\}$ from our simulations. Non-overlapping windows no larger than $10^6$ in sample size tile over the full $T_n$ process and determine how $n$ is chosen for each $\Delta T_r$. The angle brackets denote an average across every window, or equivalently over $n$ starting  values. As in the experiments \cite{Sreenivasan2002}, we find two scaling regions for each moment $q$, described by
\begin{equation}
    \langle |\Delta T_r|^q\rangle \sim r^{\zeta_q}.
       \label{eq:pl}
\end{equation}
The straight lines in Fig. \ref{fig:Tnr} show the short-lag-time scaling ($r<5$) captured by Eq. \eqref{eq:pl}, and in Fig. \ref{fig:ZetaQ} we delineate $r<5$ from the long-time scalings for $r>50$. We find the same linear dependence as in the experiments \cite{Sreenivasan2002}, with the only noticeable difference being a slightly steeper slope for our $r<5$ profile. Specifically, for $q\geq 2$ the slope is $\approx 0.52$ and the $\zeta_q$-intercept is $\approx1.38$. The linear trend for  $r>50$ is identical to that of Sreenivasan et al. \cite{Sreenivasan2002}, with a slope of 1 and the $\zeta_q$ intercept of 0. This implies the same large-$r$ decorrelation of events for $\Delta T_r$. 

{\color{black} We turn to the unique behavior of the third moment ($q=3$) in Figs. \ref{fig:Tnr} and \ref{fig:ZetaQ}.  The former shows that this generalized inter-switch 3rd moment does not deviate from the short time scaling as $r$ grows. This is reconciled by the latter plot showing agreement between the short and long time behavior for $q=3$. A broader reflection of the notion of multifractality is exhibited in Fig. 9 of the experiments of Sreenivasan et al. \cite{Sreenivasan2002} which shows $\zeta_q$ = $q$ for $q \approx 1$.
However, we note that for $r>50$, the linear $\zeta_q$ vs. $q$ relation is a hallmark of de-correlation between fluctuations and lags. This behavior is matched exactly between the experiment and our low-dimensional model. Hence, the $q=3$ moment in our numerical results simply marks a transition between the typical lower-moment fluctuations and the higher-moment extreme fluctuations. Thus it is unsurprising that for $r<5$, more typical fluctuations are favored while extreme fluctuations are heavily penalized, by the nature of stronger correlations coinciding with smaller separations.}

\begin{figure}
    \centering
    \includegraphics[width=1.0\linewidth]{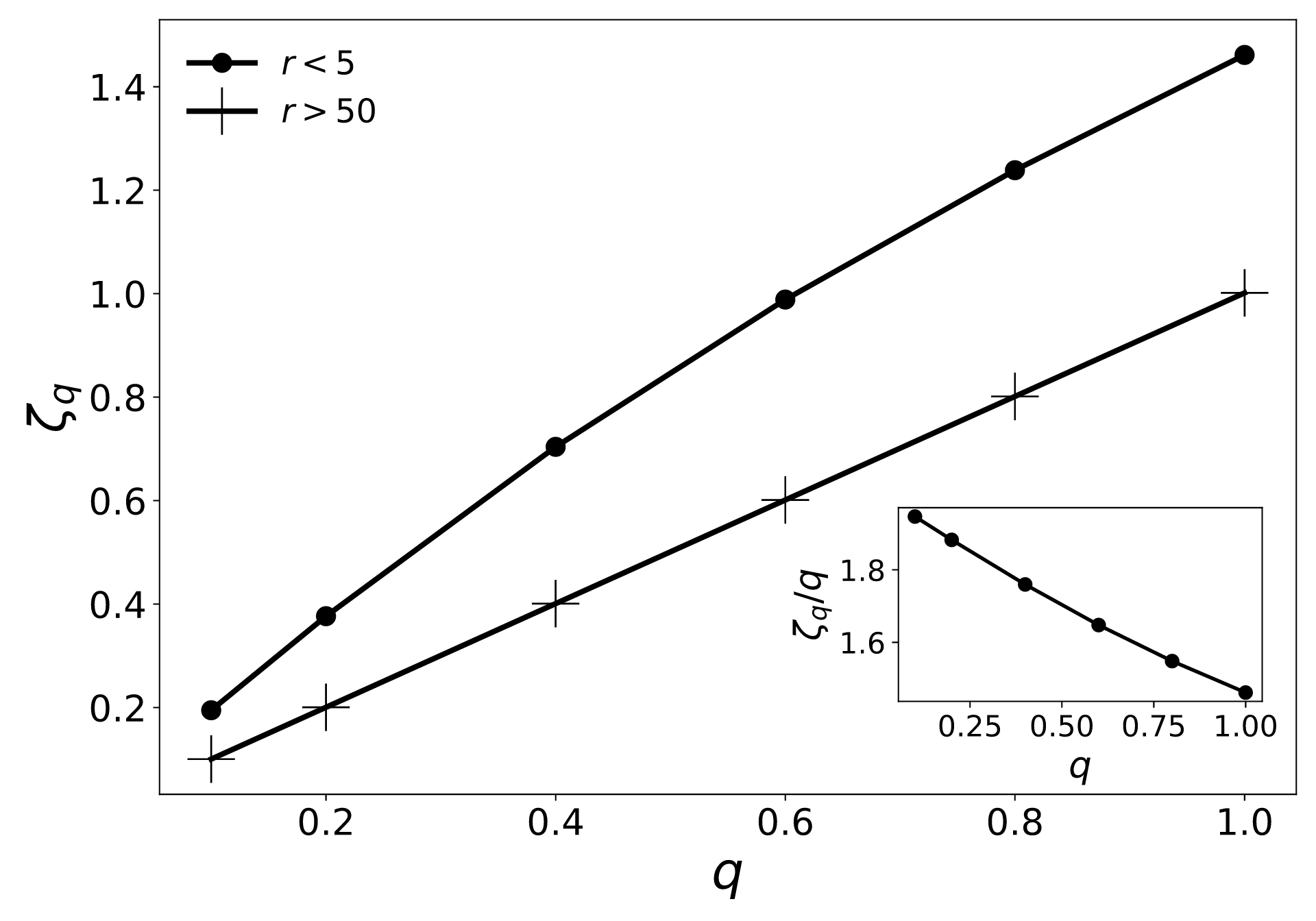}
    \caption{The local power law exponents, $\zeta_q$, are plotted against $q\leq 1$ for $r<5$ (circles) and $r >50$ (crosses). The inset displays the $q$-normalized $\zeta_q$ values against $q$.  Cf. Fig. 10 of \cite{Sreenivasan2002}.\\}
    \label{fig:zl1}
\end{figure}

In Fig. \ref{fig:zl1} we show the scaling behavior for $q\in\{0.1,0.2,0.4,0.6,0.8,1.0\}$, wherein the approximate linearity of $\zeta_q$ with $q$ for $r>50$ remains, such that the slope is 1 and the intercept is 0. Additionally, since the generalized Hurst exponent, $H(q)$, is given by
\begin{equation}
    H(q)=\zeta_q/q,
\end{equation}
the inset in Fig. \ref{fig:zl1} for $r<5$ shows that $\zeta_q$ is concave quadratic in $q$. This recovers the approximate lognormal distribution of the generalized inter-switch intervals for $r<5$.

\subsection{Multifractal Analysis}
We divide the full $G_n$ time series of length $N$ into $\lfloor N/\ell \rfloor$ non-overlapping blocks using scales $\ell$. 
The imposed measure, for some block $B_j(\ell)$, is defined by
\begin{equation}
\mu[B_j(\ell)]=\frac{\sum_{n\in B_j(\ell)} |G_n|}{\sum_{\hat{n}=1}^N|G_{\hat{n}}|},
\end{equation}
from which the $q$-moment partition function is defined by
\begin{equation}
    \chi(q,\ell)=\sum_{j}\left( \mu[B_j(\ell)] \right)^q \stackrel{\ell \to 0}{\sim} \ell^{\tau(q)}.
\end{equation}
The mass-scaling exponent, $\tau(q)$, is related to the R\'{e}nyi dimensions $D_q$ through
\begin{equation}
    (q-1)D_q=\tau(q)=\lim_{\ell\to0}\frac{\log \chi(q)}{\log \ell},
\end{equation}
and a Legendre transform \cite{Halsey} connects $\tau(q)$ to the H\"{o}lder exponent, $\alpha(q)$, and the multifractal spectrum viz.
\begin{equation}
f(\alpha):=q\alpha(q)-\tau(q).
\label{eq:legf}
\end{equation}

\begin{figure}
    \centering
    \hspace*{-7mm}
    \includegraphics[width=1.0\linewidth]{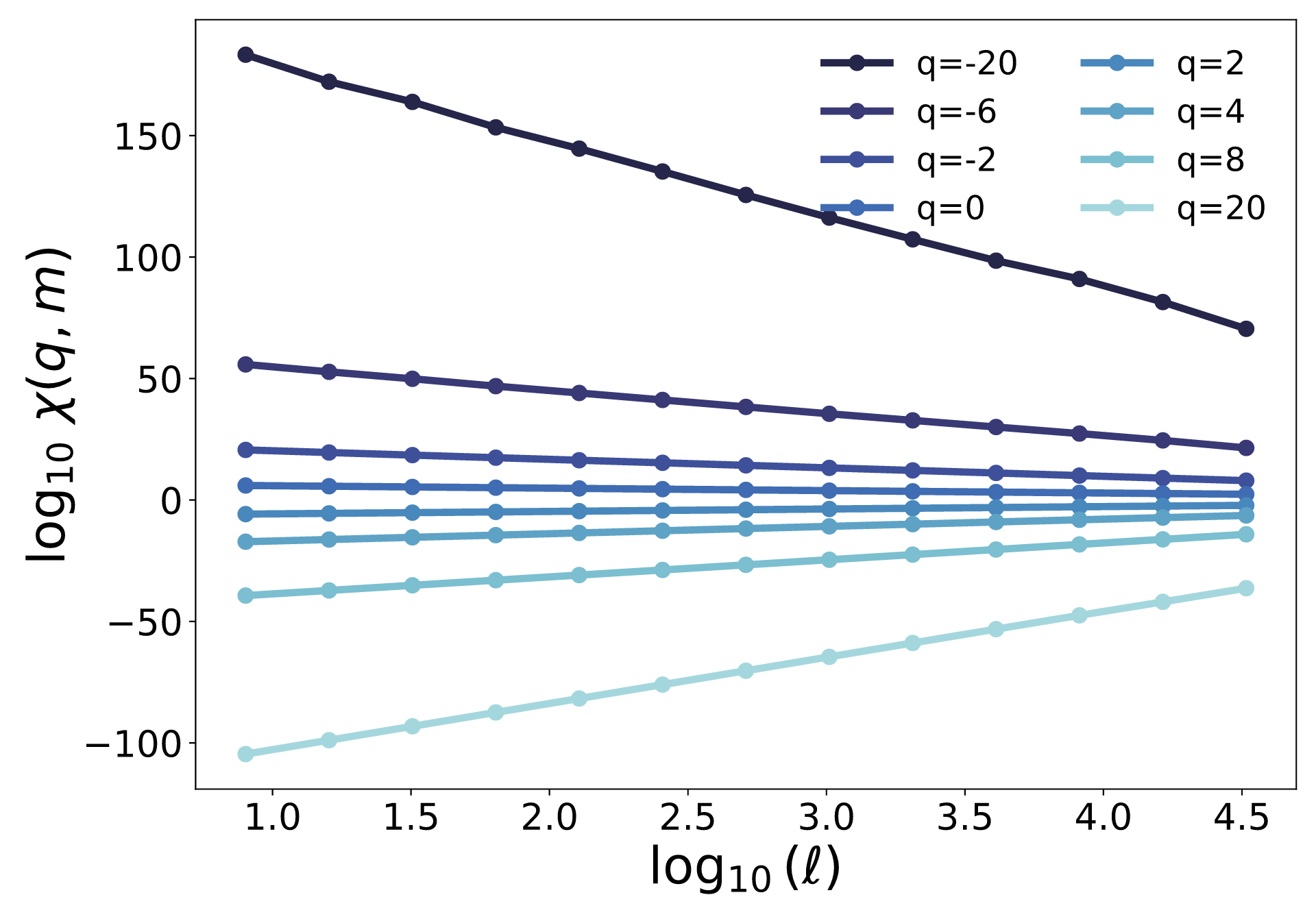}
    \caption{The $q$-moment partition sum, $\chi(q,\ell)$, is numerically computed on the $q$-grid of $q=-20,-6,-2,0,2,4,8,\text{ and } 20$ and plotted using the differently colored lines (top to bottom respectively) to display the change in the partition function's behavior as the length-scale $\ell$ coarsens.}
    \label{fig:7}
\end{figure}

\begin{figure*}
        \centering
        \includegraphics[width=0.31\linewidth]{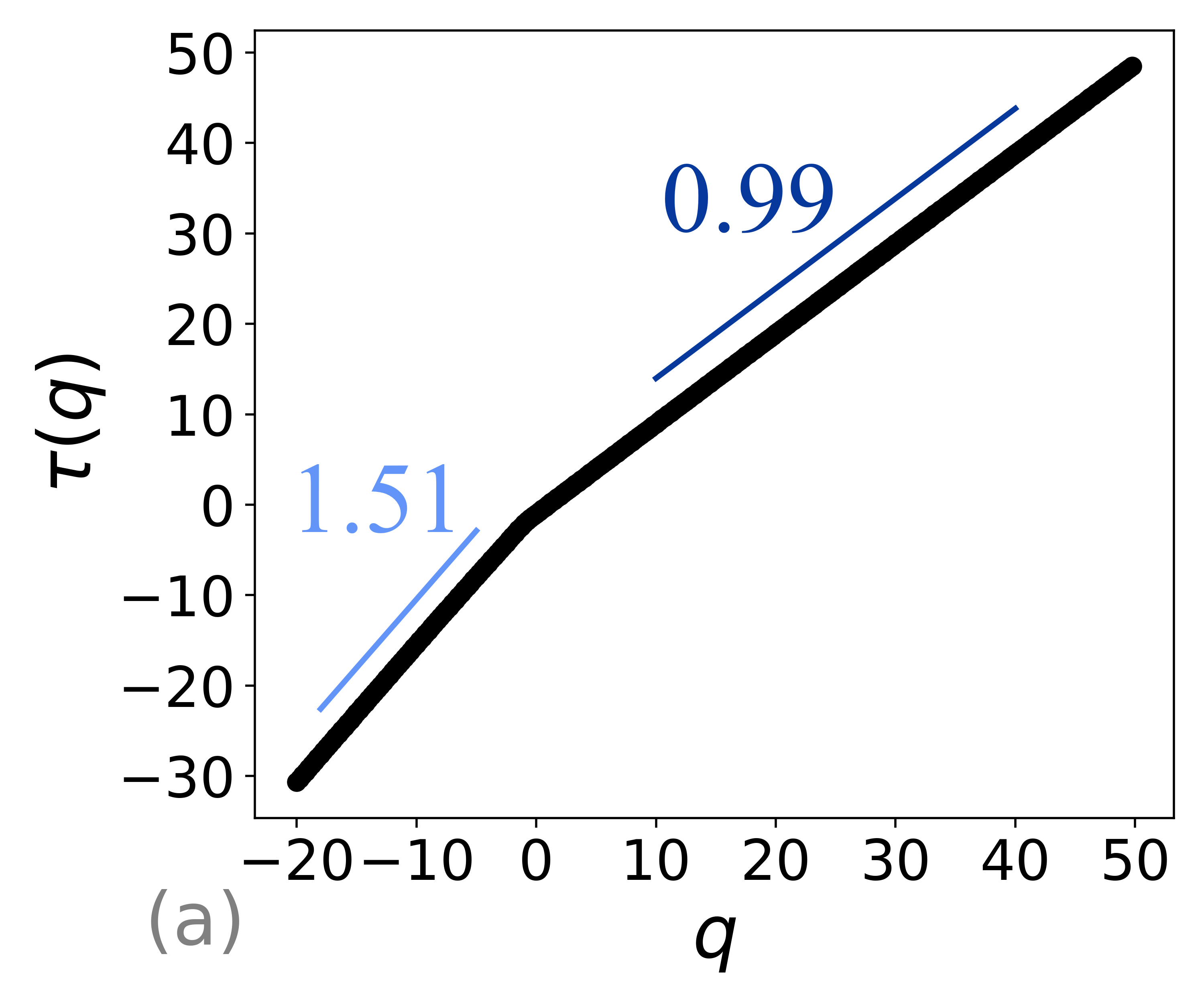}
        \includegraphics[width=0.31\linewidth]{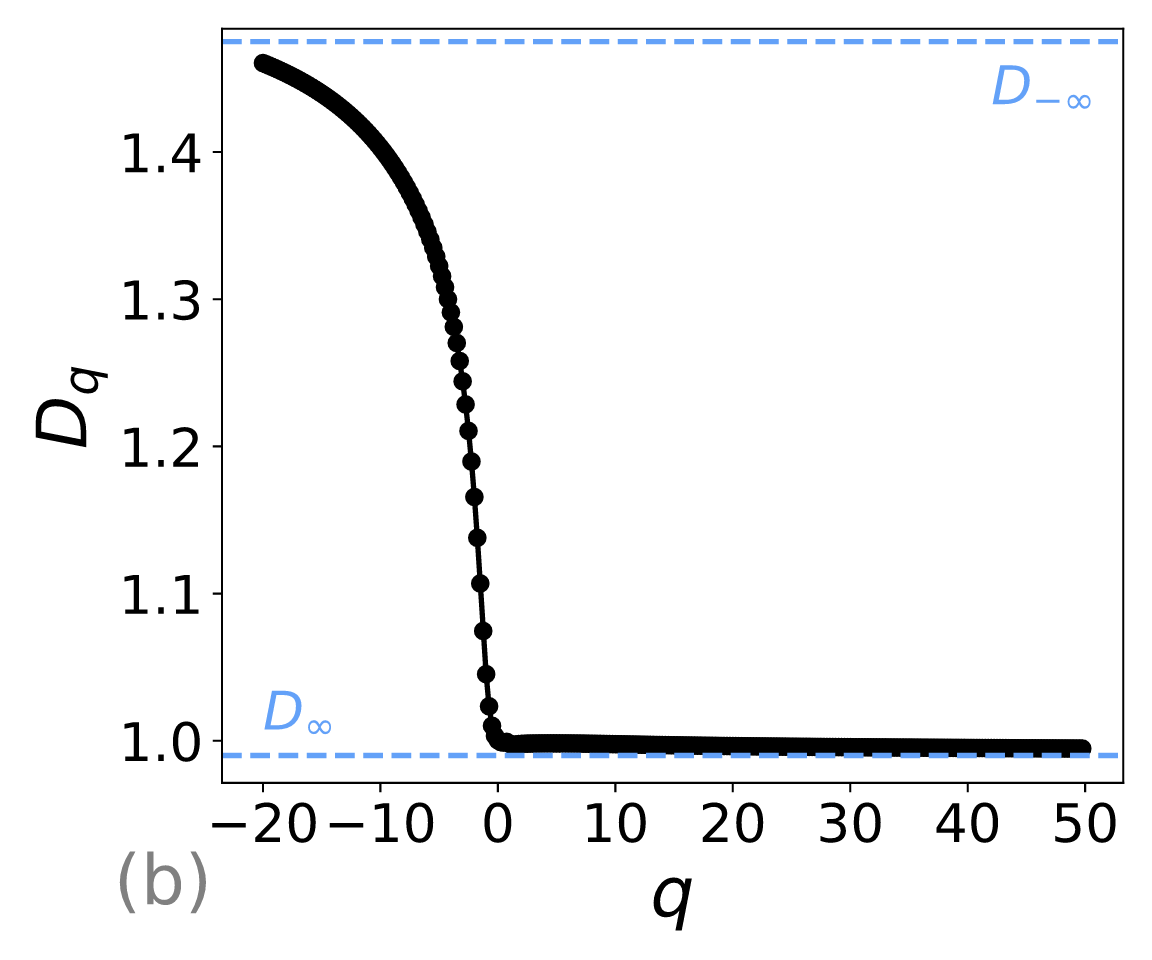}
        \includegraphics[width=0.31\linewidth]{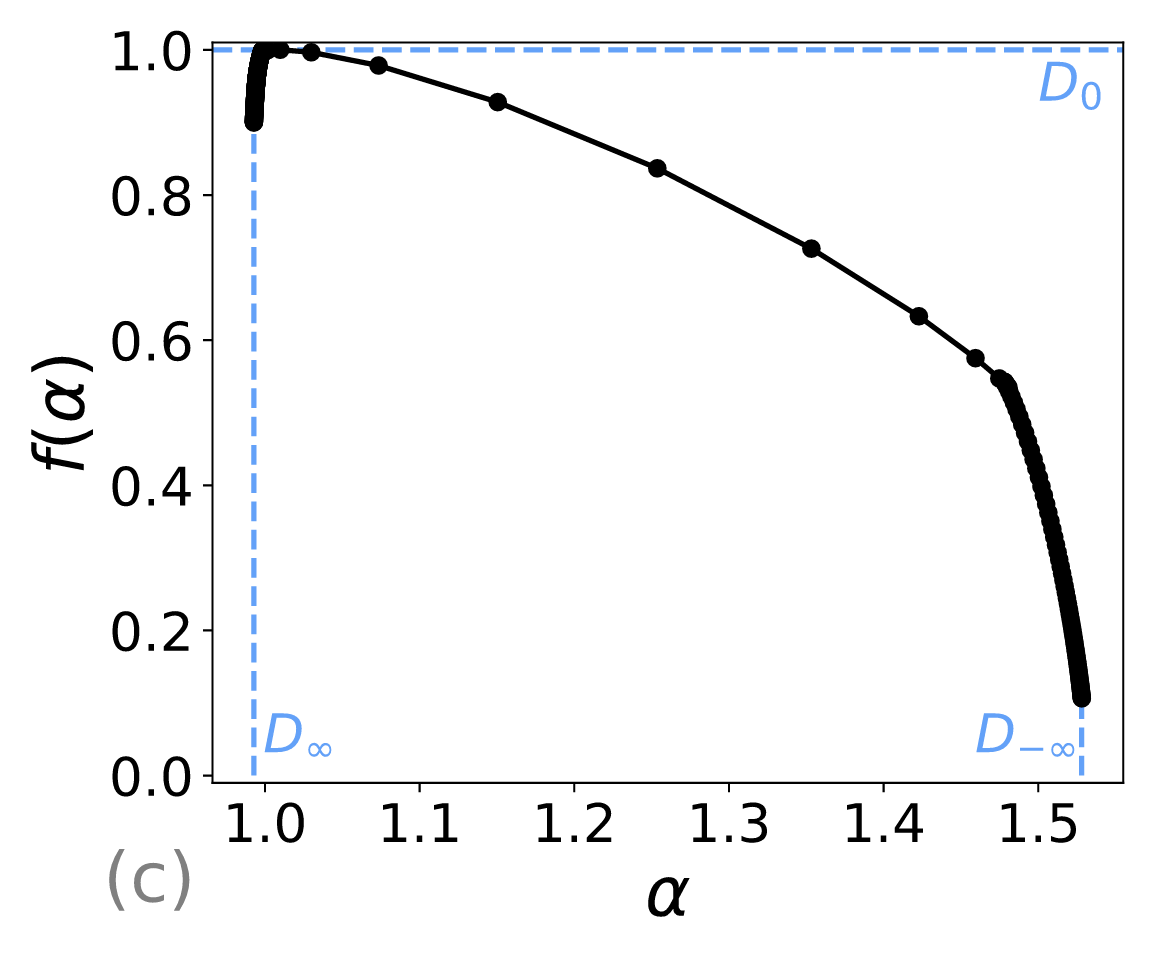}
        \caption{The numerical results for (a) the mass-scaling exponent, $\tau(q)$, plotted against each realized moment with two different slopes; (b) the R\'{e}nyi, or generalized dimensions $D_q=\tau(q)/(q-1)$, plotted against each moment $q$; and (c) the multifractal spectrum $f(\alpha)$ versus $\alpha=\alpha(q)$ for the full range of realized H\"{o}lder exponents.}
        \label{fig:MFnum}
    \end{figure*}
    
The mass-scaling exponent can be populated by the slopes of lines in Fig. \ref{fig:7}. As shown in Fig. \ref{fig:MFnum}(a), 
$\tau(q)$ has a bent spectrum, implying multifractal behavior because monofractals tend to have $\tau(q)$ non-piecewise linear in $q$ \cite{Halsey}.
We then recover the generalized dimensions $D_q$, and as shown in Fig. \ref{fig:MFnum}(b), the profile is monotonic in $q$. The box counting dimension is denoted by $D_0=1$, which is consistent with a one-dimensional support. Higher $q$-moments suggest that the largest deviations from linearity of $T_n$ vs. $n$ show a $q$-effective count of occupied boxes that grows effectively linearly. The large-magnitude $G_n$ values populate thin, 1-D sets, temporally. For $q<0$, the R\'enyi dimensions become non-integer values larger than 1. These moments populate $\chi(q,\ell)$ with the lowest magnitude oscillations of $G_n$ about zero. As $\ell$ shrinks, the growth rate of the $q$-effective count  is larger than that of the largest magnitude deviations. This indicates a much richer intermittent behavior for the smallest oscillations on the smallest time scales.

Finally, in Fig. \ref{fig:MFnum}(c) we show the multifractal spectrum computed using the values of $\tau(q)$.
The largest value of $\alpha$ in the domain of Fig. \ref{fig:MFnum}(c) corresponds to $D_{-\infty}=\alpha_{\max}=1.528$, whereas the smallest gives $D_\infty=\alpha_{\min}=0.99$. These are the extremal scaling properties of the system. The maximum of $f(\alpha)$ corresponds to $D_0=1$ as noted above. This motivates a treatment in terms of a Cantor set cascade formalism that provides an interpretive framework for these numerical results, as discussed next.

\subsection{Cantor Set Cascade}

For parsimony of notation, let the measure be $\mu[B_j(\ell)]:=p_j(\ell)$, and use a cascade that produces an asymmetric multifractal spectrum like that computed in Fig. \ref{fig:MFnum}(c). See, for example, \cite{Halsey} and the Appendix below for more details. The measure chosen, $p_j$, assigns large density to large magnitude inter-switch timing deviations and small density to the high-frequency oscillation about the linear trend. An analytic treatment of the multifractal behavior of $G_n$ begins with a Cantor set cascade with unit-length support.

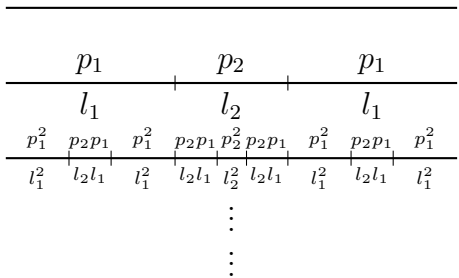
\begin{figure}
\centering
\begin{tikzpicture}[x=1cm, y=1cm, font=\large]
  \draw[thick] (0,0) -- (6,0);

  \draw[thick]
    (0,-1)   -- node[midway,above]{$p_1$} node[midway,below]{$l_1$} (2.25,-1)
             -- node[midway,above]{$p_2$} node[midway,below]{$l_2$} (3.75,-1)
             -- node[midway,above]{$p_1$} node[midway,below]{$l_1$} (6.0,-1);

  \foreach \x in {2.25,3.75} {
    \draw (\x,-1+0.1) -- (\x,-1-0.1);
  }
  \begin{scope}[font=\scriptsize]
    \draw[thick]
      (0,-2)   -- (0.84,-2) node[midway,above]{$p_1^2$} node[midway,below]{$l_1^2$}
               -- (1.4,-2) node[midway,above]{$p_2p_1$} node[midway,below]{$l_2l_1$}
               -- (2.25,-2) node[midway,above]{$p_1^2$} node[midway,below]{$l_1^2$}
      -- (2.25,-2) -- (2.81,-2) node[midway,above]{$p_2p_1$} node[midway,below]{$l_2l_1$}
               -- (3.2,-2) node[midway,above]{$p_2^2$} node[midway,below]{$l_2^2$}
               -- (3.75,-2) node[midway,above]{$p_2p_1$} node[midway,below]{$l_2l_1$}
      -- (3.75,-2) -- (4.59,-2) node[midway,above]{$p_1^2$} node[midway,below]{$l_1^2$}
               -- (5.15,-2) node[midway,above]{$p_2p_1$} node[midway,below]{$l_2l_1$}
               -- (6.0,-2) node[midway,above]{$p_1^2$} node[midway,below]{$l_1^2$};

    \foreach \x in {0.84,1.40,2.25,2.81,3.2,3.75,4.59,5.15} {
      \draw (\x,-2+0.1) -- (\x,-2-0.1);
    }
    \end{scope}
    \node at (3,-2.7) {\(\vdots\)};
    \node at (3,-3.3) {\(\vdots\)};
  \end{tikzpicture}
  \caption{A schematic for the Cantor set cascade used to recreate an analytic version of the numerically computed multifractal quantities. Masses are denoted by $p_1$ and $p_2$, while length-scales are labeled $l_1$ and $l_2$ (conditions in \eqref{eq:a(1)}).}
  \label{fig:schem}
\end{figure}

We define two equivalent branches of larger mass that correspond to large bursts in $G_n$, while a central, smaller mass branch corresponds to high-frequency scatter about the mean. Accordingly, $2p_1+p_2=1$ ($2l_1+l_2=1$), such that $p_1>p_2$ ($l_1>l_2$). {\color{black} For $n$ arbitrary cascade steps, we let $n_1$ be the number of size $l_1$ steps and $n_2$ the number of size $l_2$ steps with $n_1+n_2=n$. Such a path accumulates mass $p_n=p_1^{n_1}p_2^{n_2}$ and size $l_n=l_1^{n_1}l_2^{n_2}$. The number of such paths is $2^{n_1} {n\choose n_1}$. We then let $w:=n_1/n \in[0,1]$ and $1-w=n_2/n\in[0,1]$. Using Stirling's approximation, we deduce that
\begin{equation}
    \begin{aligned}
        {n \choose n_1} &\sim  \frac{[w^{-w}(1-w)^{-(1-w)}]^n}{\sqrt{2\pi n w (1-w)}},
    \end{aligned}
\end{equation}
where
\begin{equation}
    \begin{aligned}
        \left[w^{-w}(1-w)^{-(1-w)}\right]^n = e^{n H(w)},
    \end{aligned}
\end{equation}
with $H(w) := -w \ln w - (1-w) \ln (1-w)$. Moreover, in the large $n$ limit, $\sqrt{2\pi n w (1-w)}$ grows algebraically and thus it follows that
\begin{equation}
    {n \choose n_1}\approx e^{n H(w)}.
\end{equation}
Additionally, because $2^{n_1}=\exp(n_1 \log 2)$, the combinatorial entropy is $\sim \exp[n H(w)+nw \log 2]$ as $n_1=nw$.

The scaling argument,
\begin{equation}
N(l_n):=2^{n_1}{n\choose n_1}\sim l_n^{-f} \qquad (n \to \infty),
\end{equation}
where $N(l_n)$ is the covering number for the whole support, leads to
\begin{equation}
    f =\lim_{n\to\infty} \frac{\log \left[N(l_n) \right]}{-\log l_n}=\frac{H(w)+w \log 2}{-w \log l_1 +(1-w) \log l_2}.
\end{equation}
Solving for $\alpha$ in terms of $w$ yields
\begin{equation}
    \begin{aligned}
        \alpha(w) =\lim_{n \to \infty}\frac{\log p_n}{\log l_n}=\frac{w \log p_1 + (1-w) \log p_2}{w \log l_1 + (1-w) \log l_2}.
        \label{eq:a(w)}
    \end{aligned}
\end{equation}
Computation of $\tau(q)$ and $D_q$ follow automatically, thereby determining the three key multifractal quantities. The schematic in Fig. \ref{fig:schem} represents the iterations capturing an LP model described in \cite{Chhabra}. Namely, the singular measure we associate with our multiplicative process is constructed according to the notion of allowing each phase space ``box'' to accommodate variation in both mass and length scale. See the Appendix for an equivalent implicit Legendre formulation of the multifractal quantities.              }  

\begin{figure*}
        \centering
        \includegraphics[width=0.31\linewidth]{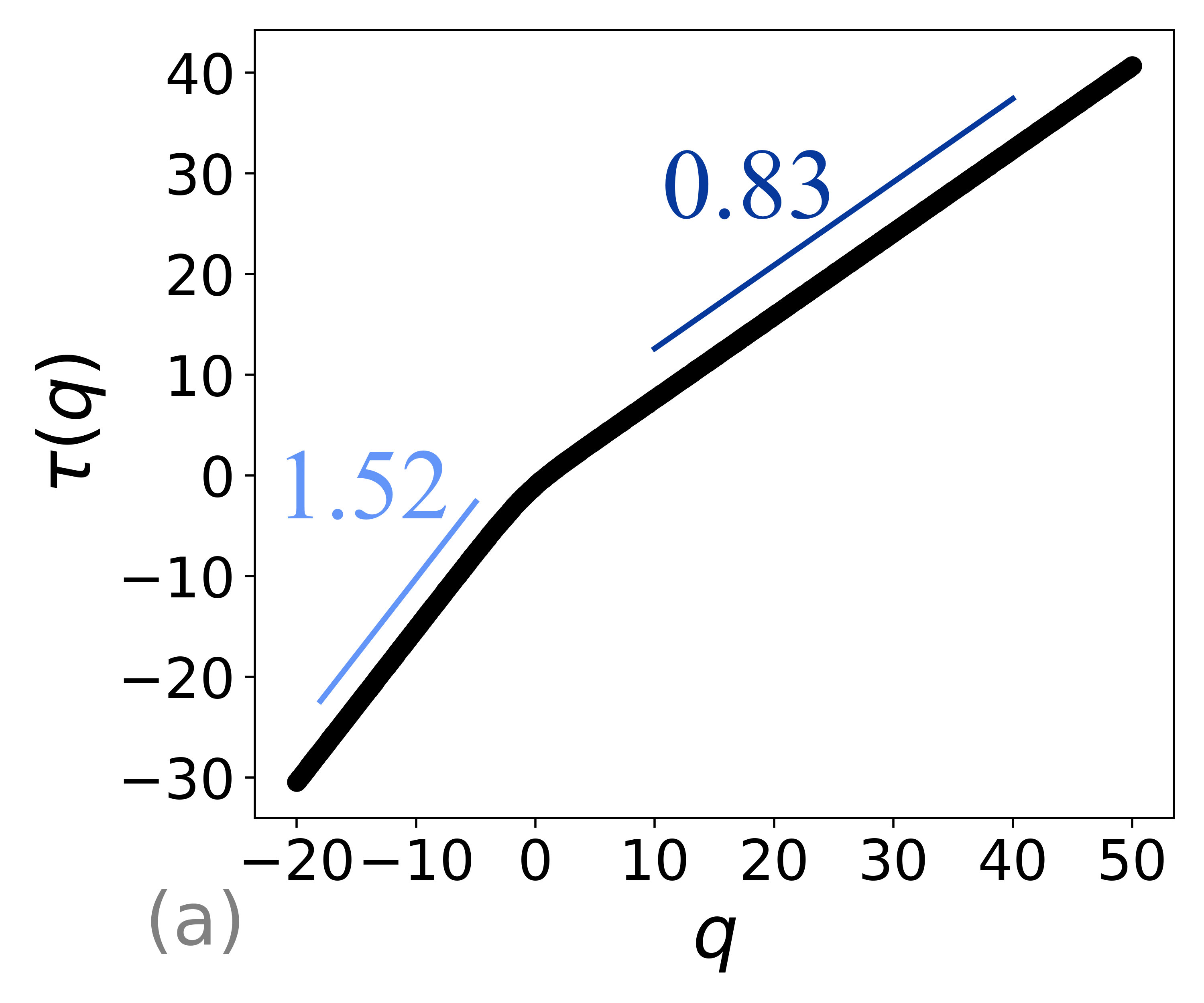}
        \includegraphics[width=0.31\linewidth]{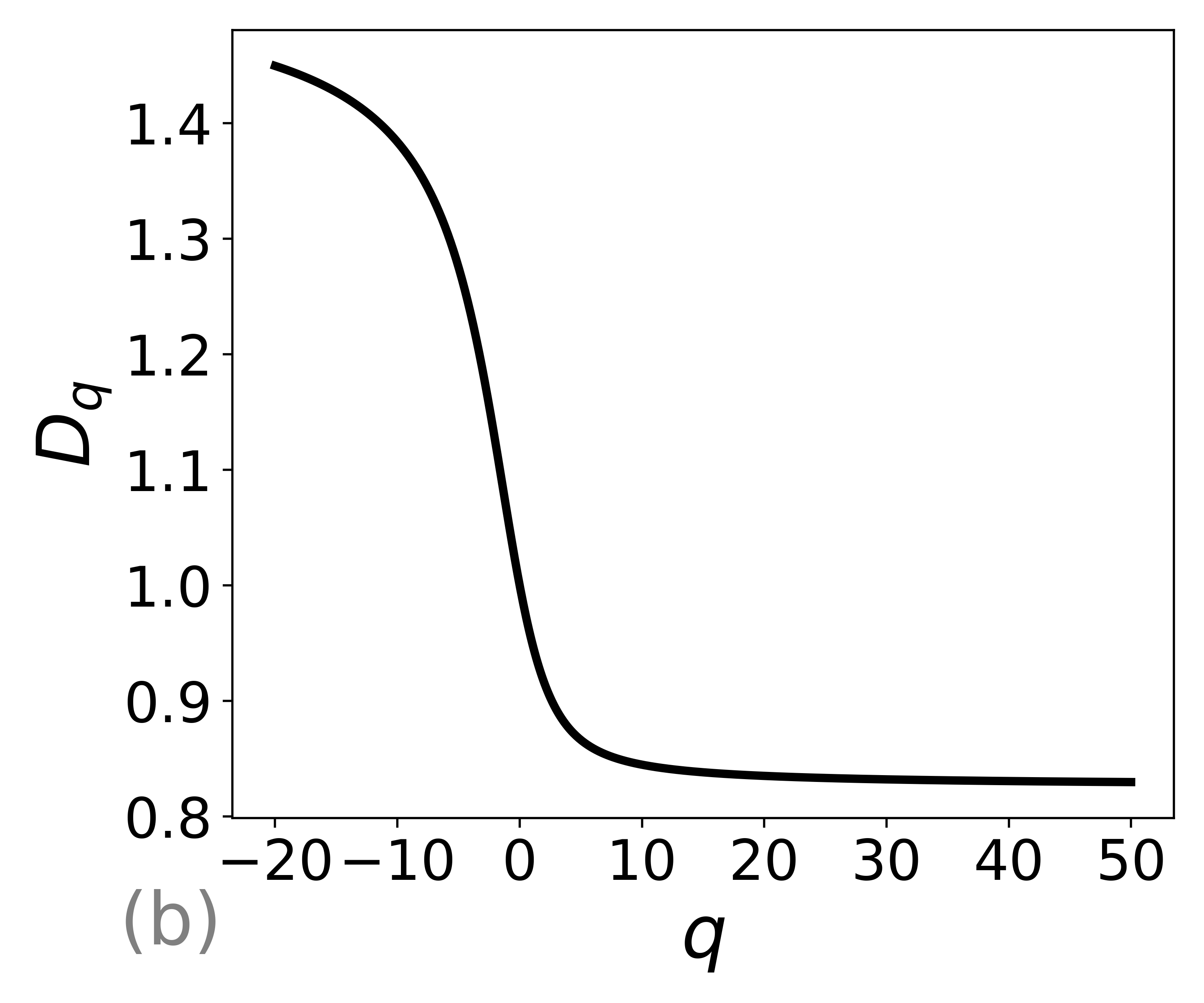}
        \includegraphics[width=0.31\linewidth]{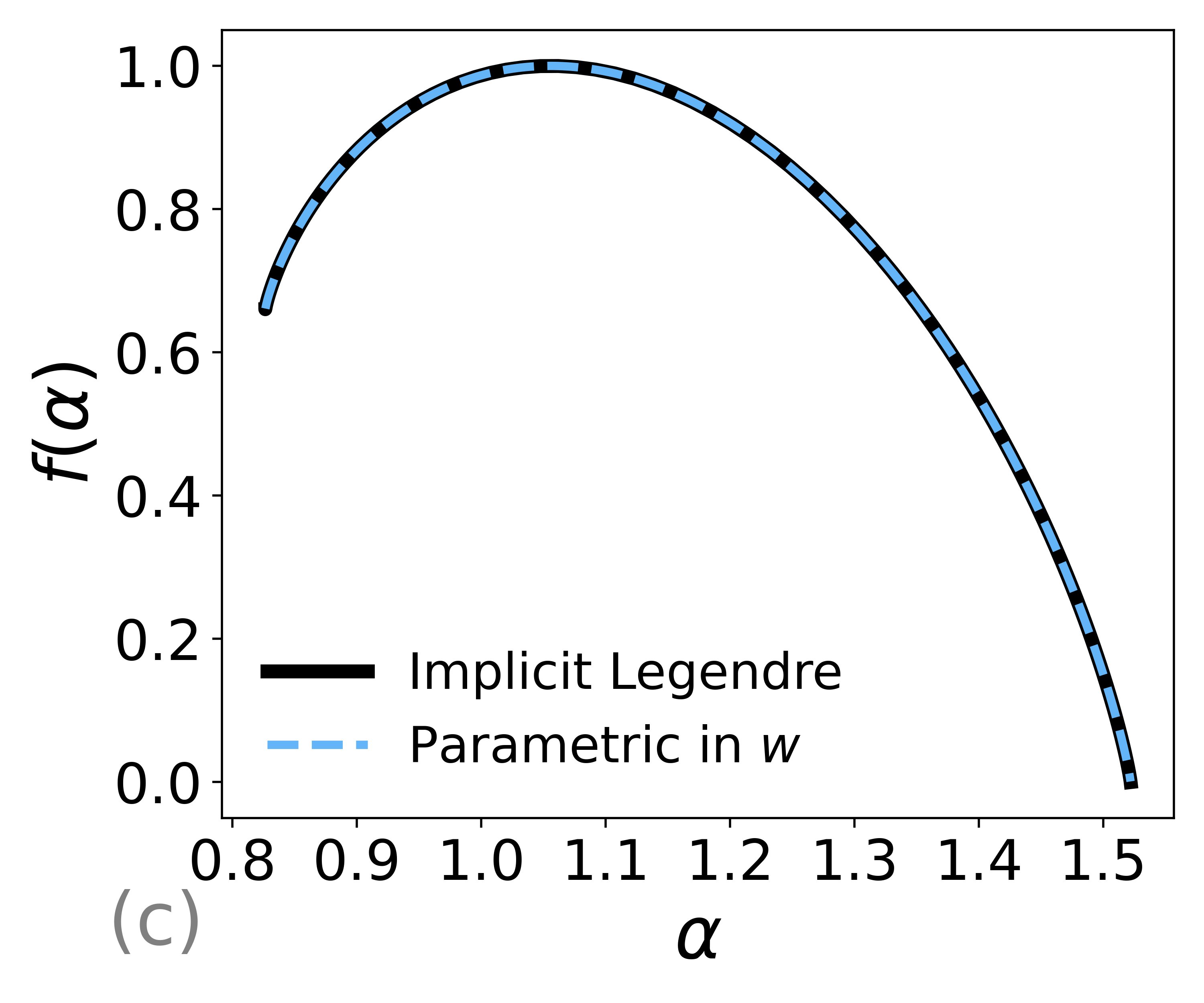}
        \caption{Analytic (a) $\tau(q)$, (b) $D_q$, and (c) $f(\alpha)$, with $l_1=0.35$, $p_1=0.42$. The analytic multifractal spectrum $f(\alpha)$ is plotted against $\alpha$ using both Eq. \eqref{eq:a(w)} in blue and Eq. \eqref{eq:a(q)} in black.}
        \label{fig:analytic}
    \end{figure*}

Fig. \ref{fig:analytic} shows that the analytic mass-scaling spectrum, R\'{e}nyi dimensions, and multifractal spectrum from the Cantor set treatment exhibit good agreement with the numerically estimated versions.

\section{Conclusion}

We have used a stochastic version of the Lorenz equations \eqref{eq:SLT} as a surrogate for the mean-wind reversals measured in the Rayleigh-B\'enard convection experiments of Sreenivasan et al. \cite{Sreenivasan2002}. In particular, we ascribed the sign changes in $X$, or the ``lobe switching'' on the stochastic Lorenz attractor, shown in Fig.~\ref{fig:1} to the experimental reversals in the mean-wind shown in Fig.~\ref{fig:a}.  The physical importance of treating noise in the $Z$-component of the Lorenz equations (the deviation of the vertical temperature profile from linearity) is to deliberately manipulate the interaction between the boundary layers and the core flow. {\color{black} This admits mathematical support by systematic analyses connecting Eqs. \ref{eq:SL} to the stochastic two-dimensional Navier-Stokes equations \cite{Zelati2021, Brz2022, Hairer2006, Stacy2026}.}

We quantified the lobe inter-switch timing as a stochastic process encoded in $G_n$ from Eq.~\eqref{eq:Gn} spanning the entire simulation record ${n\in \{1,...,N\}}$. Computing the band-limited power spectral density produced a Gaussian PDF (Fig.~\ref{fig:6}) as was found experimentally (see Fig.~4 of \cite{Sreenivasan2002}).  We suggest that the frequency range producing Gaussianity in the simulations is a scale dependent property of the convection, whereas the experimentally observed Gaussianity may be due to the intrinsic coarse-graining of the measurements, which do not fully resolve the turbulence.

We examined the second moment statistics of $G_n$ to find an autocorrelation function indicative of fractional Brownian motion, and a classical Hurst exponent of approximately $1/2$, showing the stationary increments of a Wiener processes. We found convergence of the quadratic variation to the theoretical value (Fig.~\ref{fig:15}) and showed that $G_n$ is  a lattice-sampled Wiener process in the $L^2$ sense.  While we reproduce the second order statistics of mean-wind reversals, we also find that non-linear moment scaling persists, as reflected in the non-linear multifractal spectrum as a function of the H\"older exponents. Although it is known that the deterministic Lorenz equations exhibit intermittent behavior, transitioning between stable limit cycles and strange attractors as $\rho$ changes, the addition of noise in the $Z$-component creates a cascading intermittency that triggers the scale-bridging statistics seen in high-$\mathrm{Ra}$ convection.
The fact that this stochastic Lorenz system reproduces the second order statistics of mean-wind reversals in the Rayleigh-B\'enard convection experiments of Sreenivasan et al. \cite{Sreenivasan2002} suggests that it may be valuable as a surrogate to address a range of other measurements.

\begin{acknowledgments}
The authors gratefully acknowledge 
support from the US National Science Foundation Graduate Research Fellowship Program and the Swedish Research Council under Grant No. 638-2013-9243.\\
\end{acknowledgments}
 

\appendix

\section{Cantor set scaling theory}

Here we detail the arguments surrounding Figures \ref{fig:analytic} where we employed a Cantor set theory that produces the asymmetric multifractal spectrum computed in Fig. \ref{fig:MFnum}(c).  This follows, {\em mutatis mutandis}, the generalization of the two-scale Cantor set example given by Halsey et al. \cite{Halsey}. Multifractal behavior in high-Ra thermal convection experiments was first hinted at by Wu et al. \cite{Wu}.

The weight of our measure, $p_j$, is given in terms of the contribution of $|G_n|$. The two distinct regimes delineated from the numerical values of $D_q$ are the large magnitude deviations and the small, near-zero magnitude, oscillations. Large magnitude values of $G_n$ contribute equally to $p_j$, and the near-zero oscillations contribute negligibly to the weight, requiring a cascade with two equivalent branches and one unique branch. Thus, we consider a three-branch deterministic cascade where the unit interval is divided into three segments:  two of length $l_1$, each having mass $p_1$, and one of length $l_2$, having mass $p_2$.
Therefore, we require that
\begin{equation}
    \begin{aligned}
        &2l_1+l_2=1\qquad \textrm{and} \qquad 2p_1+p_2=1 \qquad \text{where} \\
        &l_1>l_2\qquad \textrm{and} \qquad  p_1>p_2.
    \end{aligned}
    \label{eq:a(1)}
\end{equation}
Each of these three intervals are then subdivided in the same manner for $n$ such replications, as outlined by the schematic in Fig. \ref{fig:schem}. The support clearly remains continuous as it is the original, unit-length line at all iterations.

It is natural to expect that $D_0=1$ for our measure; the most dense intervals on the line segment do not contract to one point, but to a set of finite dimension. Furthermore, cascades that only visits subsequent partitions $l_1$ do so through a binary choice at every step. Thus, after $n$ such steps, there are $2^n$ intervals of size $l_1^n$. Therefore, in the large $n$ limit, these $2^n$ shrinking intervals form a Cantor-like set of Hausdorff dimension
\begin{equation}
    f(\alpha_{\min})=\lim_{n\to \infty}\frac{\log(2^n)}{-\log(l_1^n)}=\frac{\log2}{-\log l_1}>0, 
\end{equation}
so that $\alpha_{\min}=D_\infty$ yields a positive value for $f(\alpha_{\min})$.

A cascade that only visits partitions of size $l_2$ will contract to a singleton in the large $n$ limit. This means that
\begin{equation}
    f(\alpha_{\max})=\lim_{n\to \infty}\frac{\log(1^n)}{-\log(l_2^n)}=\frac{\log 1}{-\log l_2}=0,
\end{equation}
so, because $f(\alpha_{\max}=D_{-\infty})=0$, we expect agreement with Fig. \ref{fig:MFnum}(c).

This formulation then requires the partition function to be
\begin{equation}
    \Gamma(q,\tau, l_1)=2 \frac{p_1^q}{l_1^\tau}+\frac{p_2^q}{l_2^\tau}=1,
\end{equation}
where, as before, $\tau=\tau(q)=(q-1)D_q$. From Fig. \ref{fig:MFnum}(c), recall that
\begin{equation}
    \begin{aligned}
        &\alpha_{\min}=D_{\infty}=0.99, \quad f(\alpha_{\min})=0.90,\\
    &D_0=1, \quad f(1.01)=1,\\
    &\alpha_{\max}=D_{-\infty}=1.528, \quad f(\alpha_{\max})=0.\\
    \end{aligned}
\end{equation}
In order to plot an analytic $f(\alpha)$, we define $\tau$ implicitly viz.
\begin{equation}
    F(\tau,q):=\Gamma(q,\tau,l_1)-1=0.
\end{equation}
Using the implicit function theorem, we find $\alpha(q)=\frac{d\tau}{dq}$ by
\begin{equation}
    \begin{aligned}
        \frac{d\tau}{dq} =-\frac{\partial F/\partial q}{\partial F/\partial \tau},
    \end{aligned}
\end{equation}
so that 
\begin{equation}
    \alpha(q)=\frac{2p_1^ql_1^{-\tau}\log p_1 + p_2^q l_2^{-\tau}\log p_2}{2p_1^ql_1^{-\tau}\log l_1 + p_2^q l_2^{-\tau}\log l_2}.
    \label{eq:a(q)}
\end{equation}

Now, to plot these results we take $l_1=0.35$ and $p_1=0.42$. Using both Eq. \eqref{eq:a(w)} and Eq. \eqref{eq:a(q)}, Fig. \ref{fig:analytic}(c) shows the agreement between the two formulations and the analytic multifractal spectrum itself.

Fig. \ref{fig:analytic}(c) shows agreement with Fig. \ref{fig:MFnum}(c) in that $f(D_{-\infty})$ and $f(D_0)$ are approximately recovered by the analytic formulation. Additionally, the concavity and general qualitative behavior are reproduced. However, the severe asymmetry in Fig. \ref{fig:MFnum}(c) is the main disagreement between the empirical and analytic results for the dense-region asymptotics ($D_\infty$).

Finally, the computation of $D_q$ follows directly---see Fig. \ref{fig:analytic}(b). As in Fig. \ref{fig:MFnum}(b), we see that, by construction, the box-counting dimension is still 1. There is a clear depression below 1 for $D_q$ where $q>0$, which is largely a result of the severity of the asymmetry in Fig. \ref{fig:MFnum}(c) relative to that in Fig. \ref{fig:analytic}(c). Namely, it is a result of the weaker agreement between numerical and analytic results for $D_\infty$. This can be attributed to the noisiness of $G_n$ in the more densely populated regions of the phase space. Nonetheless, the qualitative comparison is encouraging; the most sparse regions correspond to larger $D_q$ when $q<0$, implying that empty space dominates. When $q>0$, $D_q$ is still relatively close to 1---the heavily-clustered points geometrically resemble a filament, which stems from the deterministic nature of the cascade and is seen empirically in Fig. \ref{fig:MFnum}(b).

\end{document}